%% file: main.tex
\documentclass[11pt]{article}
\usepackage{tikz-qtree}
\usepackage[margin=1in]{geometry}
\usepackage[normalem]{ulem}
\usepackage{amssymb,amsmath,amsfonts,eurosym,ulem,graphicx,caption,color,setspace,sectsty,comment,pdflscape, array}
\usepackage{natbib}
\usepackage[hidelinks]{hyperref}
\usepackage[bottom]{footmisc}
\usepackage{dsfont}
\usepackage{tabularx}
\usepackage{hyperref}
\usepackage{graphicx}
\usepackage{booktabs}
\usepackage{colortbl}
\usepackage{bm}
\usepackage{float}
\usepackage{multirow}
\usepackage{caption}
\usepackage{subcaption}
\usepackage{scalerel,stackengine}
\usepackage{color}
\usepackage{amssymb}
\usepackage[bottom]{footmisc}
\usepackage{caption}
\usepackage{amsmath}
\usepackage{amsthm}
\usepackage{epsfig}
\usepackage{dsfont}
\usepackage{mathtools}
\usepackage{booktabs}
\usepackage{graphicx}
\usepackage{graphics}
\usepackage{float}
\usepackage{multirow}
\usepackage{color}
\usepackage{caption}
\usepackage{enumerate}
\usepackage{subcaption}
\usepackage{fullpage}
\usepackage{tikz-qtree}
\usetikzlibrary{positioning}
\usepackage{makeidx}
\usepackage{xspace}
\usepackage{wrapfig}
\usepackage{tikz}
\usepackage{natbib}
\usepackage{array,makecell}
\usepackage{multirow}
\usepackage{algorithm}
\usepackage{algpseudocode}
\usepackage[resetlabels]{multibib}
\newcites{App}{References}

\usepackage[affil-it]{authblk}

\numberwithin{equation}{section}

\newtheorem{proposition}{Proposition}
\newtheorem{Assumption}{Assumption}

\makeatletter
\newtheorem*{rep@theorem}{\rep@title}
\newcommand{\newreptheorem}[2]{%
\newenvironment{rep#1}[1]{%
 \def\rep@title{#2 \ref{##1}}%
 \begin{rep@theorem}}%
 {\end{rep@theorem}}}
\makeatother

\newcommand{\pr}{\text{P}}

\newcommand{\E}{\mathbb{E}}

\newcommand{\bX}{\mathbf{X}}
\newcommand{\bx}{\mathbf{x}}

\newlength\myindent
\usepackage{hyperref}
\hypersetup{
colorlinks=true,
citecolor=blue,
linkcolor=black,
filecolor=blue,      
urlcolor=black,
}
\begin{document}

\begin{titlepage}
\title{Causal Rule Ensemble: Interpretable Discovery and Inference of Heterogeneous Treatment Effects
\footnotetext{\scriptsize We are grateful for helpful feedback from organizers and participants at UCLA, Stanford University, Harvard University, UCSF, New York University, ETH, École Polytechnique Fédérale de Lausanne, George Washington University, St. Gallen University, KU Leuven, IMT School for Advanced Studies seminars, at the Online Causal Inference seminar (OCIS), at the IMS International Conference on Statistics and Data Science (ICSDS), at the Health Econometrics Workshop at Emory University, at the Data Science for Impact Evaluation Webinar (DSIE), and the European Causal Inference Meeting (EUROCIM). This work was supported by EPA grant (83587201-0), NIH grants (R01ES026217, R01MD012769, R01ES028033, 1R01AG060232-01A1, 1RF1AG074372-01A1, 1RF1AG080948, 1R01ES030616, 1R01AG066793, 1R01AG066793-01R01, 1R01ES029950) as well as the Alfred P. Sloan Foundation Grant G-2020-13946 for the development of ``Causal Inference with Complex Treatment Regimes: Design, Identification, Estimation, and Heterogeneity'', the VPR - Harvard University grant on ``Climate Change Solutions Fund'', the Harvard Data Science Initiative Postdoctoral Research Fund Award and the National Research Foundation of Korea (NRF) grant funded by the Korea government (MSIT) (2021R1C1C1012750).}}
\author[1]{\normalsize Falco J. Bargagli-Stoffi \thanks{Co-first authors.}}
\author[2]{\normalsize Riccardo Cadei*}
\author[1]{\normalsize Lauren Mock}
\author[3]{\normalsize Kwonsang Lee \thanks{Corresponding author. \url{kwonsanglee@snu.ac.kr}}}
\author[1]{\normalsize Francesca Dominici}
\affil[1]{Department of Biostatistics, Harvard T.H. Chan School of Public Health}
\affil[2]{Department of Computer Science, Institute of Science and Technology Austria}
\affil[3]{Department of Statistics, Seoul National University}
\date{}
\maketitle
\vspace{-1cm}
\begin{abstract}
\noindent{In health and social sciences, it is critically important to identify subgroups of the study population where there is notable heterogeneity of treatment effects (HTE) with respect to the population average. Decision trees have been proposed and commonly adopted for the data-driven discovery of HTE due to their high level of interpretability. However, single-tree discovery of HTE can be unstable and oversimplified. This paper introduces the Causal Rule Ensemble (CRE), a new method for HTE discovery and estimation using an ensemble-of-trees approach. CRE offers several key features, including 1) an interpretable representation of the HTE; 2) the ability to explore complex heterogeneity patterns; and 3) high stability in subgroups discovery. The discovered subgroups are defined in terms of interpretable decision rules. Estimation of subgroup-specific causal effects is performed via a two-stage approach, for which we provide theoretical guarantees. Through simulations, we show that the CRE method is highly competitive compared to state-of-the-art techniques. Finally, we apply CRE to discover the heterogeneous health effects of exposure to air pollution on mortality for 35.3 million Medicare beneficiaries across the contiguous U.S.}
\bigskip
\vspace{0in}
\vspace{0in}\\
\end{abstract}
\thispagestyle{empty}
\end{titlepage}

\newpage

\pagenumbering{arabic}
\setcounter{page}{1}

\begin{doublespace}

\input{1_introduction}

\input{2_setup}

\input{3_method}

\input{4_simulations}

\input{5_application}

\input{6_conclusion}

\end{doublespace}

\begin{singlespace}
\small
\bibliography{biblio}
\end{singlespace}
\pagebreak

\appendix
\counterwithin{figure}{section}
\counterwithin{table}{section}
 \pagenumbering{arabic}
    \setcounter{page}{1}

\begin{center}
    \textbf{\large SUPPLEMENTARY MATERIAL TO\\ ``CAUSAL RULE ENSEMBLE: INTERPRETABLE DISCOVERY AND INFERENCE OF HETEROGENEOUS TREATMENT EFFECTS''}\\ \vspace{0.25cm}
    \normalsize FALCO J. BARGAGLI STOFFI, RICCARDO CADEI, LAUREN MOCK, KWONSANG LEE, AND FRANCESCA DOMINICI
\end{center}

\begin{doublespace}
\input{Appendix/appendix_a}

\input{Appendix/appendix_a1}

\input{Appendix/appendix_b}

\input{Appendix/appendix_c}

\input{Appendix/appendix_d}

\input{Appendix/appendix_f}

\input{Appendix/appendix_g}

\input{Appendix/appendix_e}
\end{doublespace}

\begin{singlespace}
\small
\bibliographystyleApp{dcu}
\bibliographyApp{biblio}
\end{singlespace}

\end{document}

%% file: 1_introduction.tex
\section{Introduction}
\label{intro}

\subsection{Motivating Application}

The U.S. Environmental Protection Agency (EPA) has recently set the goal of achieving environmental justice by addressing the disproportionate vulnerabilities in adverse human health effects due to exposure to air pollution \citep{epa2022a}. According to the EPA, environmental justice is defined as ``\textit{no group of people should bear a disproportionate burden of environmental harms and risks}'' \cite[see][page 5673]{epa2022b}. In an effort to promote environmental justice, the EPA has called for scientific studies that would inform the understanding of the disproportionate health impacts of air pollution, with particular attention to demographic-specific information \citep{epa2022b}. Despite strong evidence that exposure to air pollution increases the risk of mortality and morbidity \cite[see, e.g.,][]{schwartz2021national, wu2020evaluating, nethery2020evaluation, carone2020pursuit}, little is known about which are the subgroups---i.e., subsets of the population characterized by a given covariate-profile (e.g., female individuals, low-income \& male individuals)---who are most vulnerable or resilient to exposure to higher levels of air pollution. 

Previous air pollution vulnerability studies are limited in scope. For example, \cite{lee2021discovering} and \cite{zorzetto2024confounder} recently proposed machine learning and Bayesian nonparametric methods to assess the heterogeneity of causal effects of air pollution exposure on health outcomes. However, these studies are limited by the computational scalability of the employed methods, and their geographic coverage is restricted to selected areas of the U.S.---i.e., New England and Texas. \cite{di2017air} estimated associations between long-term exposure to air pollution and mortality rates for prespecified population subgroups defined by age, gender, and race categories. Despite its national coverage, this study has the main limitations of not directly answering a causal question, but an associational one, and providing a very limited heterogeneity exploration---i.e., heterogeneous associations are estimated just for a predefined and very limited set of characteristics (e.g., sex, age, race). Thus, despite the urgency of extensive exploration of the heterogeneous health effects of air pollution, a national study on this topic that is also based on methods of causal inference is not yet available.

The goal of our motivating application is to provide nationwide data-driven characterization and quantification of the heterogeneity of causal effects of ambient exposure to air pollution on health outcome. In particular, we develop a new interpretable causal machine learning method with the objective of identifying \textit{de novo} which subgroups of the Medicare population are most vulnerable or resilient to long-term exposure to fine particulate matter---known as PM$_{2.5}$---on mortality. To do so, we acquired and integrated the data on 35,331,290 Medicare beneficiaries (i.e., individuals 65 years of age or older) in the contiguous U.S. for the period 2010-2016. We consider a binary exposure, indicating whether each individual has been exposed to PM$_{2.5}$ greater than 12 $\mu g \slash m^3$ or not. This binary exposure is extremely relevant from a policy perspective as it was the National Ambient Air Quality Standard (NAAQS) established by the EPA during this period of time. We link exposure to two-year annual PM$_{2.5}$ during 2010-2011 in the zip code of residence to an individual-level indicator of mortality during the 5-year period 2012-2016 and several potential confounders at the individual, zip code, and county levels. Our study focuses on characterizing heterogeneity in the causal effects within the four geographic regions of the U.S. Census Bureau---namely, Northeast, Midwest, West, and South.%More details about the study design and results are illustrated in section \ref{sec:application}. 

\subsection{Contribution and Related Works} \label{sec:reference}

The bulk of the heterogeneous treatment effect (HTE) literature focuses on two major tasks \citep{dwivedi2020stable}: (i) estimating HTEs (e.g., estimating the conditional average treatment effect (CATE)); (ii) discovering the set of characteristics that lead to subgroups of a population having HTE notably different from to the population average. 

Seminal work on CATE estimation is based on nearest-neighbor matching and kernel methods \citep{crump2008nonparametric, lee2009non}. Other nonparametric machine learning methods such as the random forest \citep{breiman2001random} and the Bayesian additive regression tree (BART) \citep{chipman2010bart} have been extended to estimate HTEs---see, e.g., \cite{foster2011subgroup}, \cite{hill2011bayesian} and \cite{hahn2020bayesian}. \cite{wager2018estimation} and \cite{athey2019generalized} developed forest-based methods for the estimation of HTE. They also provide an asymptotic theory for the conditional treatment effect estimators and valid statistical inference. Recently, two-stage doubly robust CATE estimators have been proposed to first generate doubly robust pseudo-outcomes and then regress them onto an a priori defined set of effect modifiers \citep{kennedy2020optimal, semenova2021debiased}.

Various methodologies have also been proposed to identify the subgroups that characterize HTE \citep{imai2013estimating, qian2011performance, Kennedy2017, Nie2017, crabbe2022benchmarking}. Some methods first estimate the CATE as a function of some set of covariates and then identify heterogeneous subgroups in a second stage \citep{foster2011subgroup, bargagli2020heterogeneous, hahn2020bayesian, bargagli2022heterogeneous}. Another approach is the direct data-driven discovery of heterogeneous subgroups \citep{wang2022causal, nagpal2020interpretable,wan2023rule}. Many of the methodologies in this category are decision tree-based methodologies \cite[see, e.g.,][]{athey2016, bargagli2020causal, lee2021discovering, yang2021, bargagli2020heterogeneous, bargagli2022heterogeneous}. Tree-based approaches have been widely adopted for HTE due to their appealing features. In fact, these methods are based on efficient and easily implementable recursive mathematical programming, they can be easily tweaked and adapted to different scenarios on the basis of the research question of interest, and they guarantee a high degree of interpretability.

Despite these attractive features, the discovery of single-tree heterogeneity is characterized by two main limitations: (i) instability in the identification of subgroups and (ii) reduced exploration of the potential heterogeneity---i.e., exploration of the HTE for a limited number of subgroups, namely, the set nodes of a single tree. Firstly, single-tree-based subgroup identification is sensitive to variations in the training sample---e.g., if the data are slightly altered, a completely different set of discovered subgroups might be found \cite[see, e.g.,][for discussions of this issue]{breiman1996heuristics, hastie2009elements, kuhn2013applied, bertsimas2023improving}. Secondly, it may fail to explore a vast number of potential subgroups (limited subgroup exploration)---e.g., the subgroups discovered are only those that can be represented by a single tree \citep{kuhn2013applied, spanbauer2021nonparametric}. To illustrate, consider a scenario in which two distinct factors contribute independently to the heterogeneity in treatment effects. In such cases, a single-tree algorithm may detect only one of these factors, failing to identify the second. In instances where both factors are identified by the tree, they are detected suboptimally as an interaction between the two variables rather than as distinct drivers of the treatment heterogeneity.

To account for these shortcomings, we propose a novel method, the Causal Rule Ensemble (CRE). CRE uses multiple trees, rather than a single tree, to discover key drivers of heterogeneity in the treatment effect via decision rules in a data-driven way. CRE provides (i) an interpretable representation of the HTE, (ii) an extensive exploration of complex heterogeneity patterns, and (iii) a guarantee of high stability in the discovery. We also develop a general two-stage estimation approach for the conditional causal effects of the discovered subgroups and provide theoretical guarantees.

CRE ensures interpretability by providing a linear decomposition of the HTE in terms of \textit{decision rules}. Interpretability is a non-mathematical concept, yet it is often defined as the degree to which a human can understand the cause of a decision \citep{kim2016examples, miller2018explanation, lakkaraju2016interpretable, wang2022causal}. \textit{If-then} decision rules are highly interpretable as they resemble human decision-making processes. The discovery of these decision rules is obtained via an extensive exploration of complex heterogeneity patterns. In particular, CRE generates candidate decision rules from an ensemble of decision trees extracting heterogeneity in the treatment effect. Among these candidate decision rules, CRE extracts only a stable set of decision rules that characterize the HTE by a rework of the stability selection algorithm \citep{meinshausen2010stability}. The stability of statistical results relative to \textit{reasonable perturbations} to data and to the model used is critically important for reproducible research \citep{yu2013stability}. In addition to enhanced reproducibility, the stability selection algorithm also allows controlling for finite sample false discovery error.

Finally, CRE provides a two-stage estimation approach for the estimation of the coefficients in the discovered linear model of the CATE. In the first stage, pseudo-outcomes are produced using any of the available techniques for the estimation of HTE at the individual level. In the second stage, these pseudo-outcomes are regressed onto the discovered rules. Different subsamples are used for the discovery of decision rules and for the estimation of HTE to avoid overfitting (that is, honest splitting by \cite{athey2016}). We provide theoretical results that guarantee the consistency and asymptotic normality of the estimated model coefficients. %We also note that the proposed two-stage estimation is similar in spirit (even if the target estimands are different) to the doubly robust (DR) learner proposed by \cite{kennedy2020optimal}. %A new sensitivity analysis is finally developed for the AATE estimates to assess their robustness to potential unmeasured confounding bias.

The remainder of the paper is organized as follows. In Section \ref{sec:formulation}, we introduce the causal setup and present a novel interpretable characterization of HTE via decision rules. In Section \ref{sec:method}, we introduce the proposed CRE methodology. In Section \ref{sec:simulations}, we validate this methodology by simulated experiments, which are further extended in the Supplementary Material. In Section \ref{sec:application}, we propose to answer the EPA's call for environmental justice, applying the CRE method to assess vulnerability and resilience from air pollution exposure in the U.S. Section \ref{sec:conclusion} discusses the strengths and weaknesses of our proposed approach and the areas of future research. CRE is implemented in an \texttt{R} package available on \href{https://cran.r-project.org/web/packages/CRE/index.html}{\texttt{CRAN}}. The complete documentation for the package can be found at \href{https://nsaph-software.github.io/CRE/}{\texttt{https://nsaph-software.github.io/CRE/}}. The code for the simulations and the application can be found at \href{https://github.com/NSAPH-Projects/cre\_applications}{\texttt{https://github.com/NSAPH-Projects/cre\_applications}}.

%% file: 2_setup.tex
\section{Problem Formulation} 
\label{sec:formulation}

\subsection{Causal Setup}
\label{ssec:potentialputcome}

Let $\mathcal{I}$ be a sample of $N$ individuals. For each individual $i\in \mathcal{I}$, let $\bm{X}_i \in \mathcal{X} \subseteq \mathbb{R}^P$ the set of observed covariates, $Z_i \in \{0,1\}$ the observed (binary) treatment, and $Y_i \in \mathcal{Y} \subseteq \mathbb{R}$ the observed outcome. Following the potential outcome framework \citep{rubin1974estimating}, for each individual, we define $Y_i(1)$ and $Y_i(0)$ as the potential outcomes under treatment and control, respectively; and the Individual Treatment Effect (ITE), $\tau_i:= Y_i(1) - Y_i(0).$ The Average Treatment Effect (ATE) is the expected value of the ITE, $\bar{\tau}:= \mathbb{E} \left[Y_i(1) - Y_i(0) \right].$ The Conditional Average Treatment Effects (CATE) on $S \subseteq \mathcal{X}$, is the expected value of the ITE conditioning over a subset of the covariates space:
\begin{equation}\label{eq:gate}
    \tau(S)  := \mathbb{E} \left[Y_i(1) - Y_i(0) \mid \bm{X}_i \in S \right].
\end{equation}
The CATE can be specified at different levels of \textit{granularity} depending on the complexity of $S$. For example, at the highest level of granularity, one might want to estimate the Individual Average Treatment Effect (IATE): $\tau(\bm{x}):= \mathbb{E} \left[Y_i(1) - Y_i(0) \mid \bm{X}_i = \bm{x} \right]$. At a lower level of granularity, one might want to estimate the average treatment effect for some \textit{subgroups} of the population. This latter estimand can also be referred to as the Group Average Treatment Effect (GATE) \citep{jacob2019group}. Throughout this paper, we will use the more general term of CATE. In fact, both IATE and GATE can be seen as special cases of CATE.

Since only one potential outcome can be observed for each individual---the fundamental problem of causal inference \citep{holland1986statistics}---we need to rely on a few assumptions to identify the causal estimands of interest.
\begin{Assumption}[Stable Unit Treatment Value Assumption (SUTVA)]
\label{ass:SUTVA}
	\begin{equation*}
		\begin{aligned}
			(i). \:\:\: Y_i(Z_i) &= Y_i, \qquad \forall i \in \mathcal{I}\\
			(ii). \:\:\:Y_i(Z_i) &= Y_i(Z_1, Z_2, \cdots, Z_i, \cdots,  Z_{N}) \qquad \forall i \in \mathcal{I}.
		\end{aligned}    
	\end{equation*}
\end{Assumption}
\noindent SUTVA enforces that for each individual $i$, $i$'s outcome is simply a function of $i$'s treatment. This is a combination of (i) consistency (no different versions of the treatment levels assigned to each unit) and (ii) no interference assumption (among the individuals) \citep{rubin1986comment}.

\begin{Assumption}[Strong Ignorability]
\label{ass:ignorability}
Given the observed covariate vector $x_i$, the treatment assignment is strongly ignorable if
\begin{equation*}
   \{Y_i(1), Y_i(0)\} \perp Z_i \mid \bm{X}_i = \bm{x}_i, 
\end{equation*}
and $0<Pr(Z_i=1 \mid \bm{X}_i=\bm{x}_i)<1$, for  all $i=1, \dots, n$.
\end{Assumption}
This assumption states that (i) the two potential outcomes are independent of the treatment conditioning on some covariate values, and (ii) all units have a strictly positive chance of receiving or not receiving the treatment.

Under Assumptions \ref{ass:SUTVA}, and \ref{ass:ignorability}, the CATE can be identified as:
\begin{equation}
\label{eq:CATEidentification}
   \tau(S)  = \mathbb{E} \left[Y_i \mid \bm{X}_i \in S, Z_i = 1 \right] - \mathbb{E} \left[Y_i \mid \bm{X}_i \in S, Z_i = 0 \right].
\end{equation}
%It is uncertain whether the set of covariates taken into consideration is adequate for establishing unconfoundedness. If it is not the case, the identification results do not hold. Sensitivity analysis provides a useful tool to investigate the impact of unmeasured confounding bias.
Several algorithms have already been proposed for the estimation of CATE under the assumptions mentioned above \cite[see, e.g.,][]{hill2011bayesian, foster2011subgroup, wager2018estimation, hahn2020bayesian, athey2019generalized}. Although they can estimate the CATE accurately, these algorithms are not built directly to provide an interpretable characterization of the HTE. To account for this shortcoming, we propose a new CATE characterization in terms of easily interpretable decision rules.

\subsection{Characterizing CATE via Decision Rules}
\label{ssec:heterogeneity}

A decision rule $r \colon \mathcal{X} \to \{0,1\}$ is a general function of the covariate space $\mathcal{X}$ characterizing a specific subgroup $S \subseteq \mathcal{X}$. We particularly focus on (interpretable) decision rules whose supports decompose as $S=S_1 \times \cdots \times S_P$, where $S_p \subseteq \mathbb{R}$ for all $p \in \{1, ..., P\}$. In the remainder of the paper, we use the term \textit{decision rule} referring to this specific definition:
\begin{equation}
 r(\bm{x}):= \prod_{p=1}^P \mathds{1}(x_p \in S_{p}).
\end{equation}
In Figure \ref{fig:example_tree}, we report a dummy (binary) decision tree to provide toy examples of decision rules with two binary covariates, $x_F$ (for female) and $x_Y$ (for young). Indeed, each node in a decision tree (with the exclusion of the root), combining the conditions of all its ancestors, aligns with the above definition of decision rule. For instance, the young female subgroup is expressed by $r_4(\bm{x}) = \mathds{1}(x_F = 1) \cdot \mathds{1}( x_Y = 1)$; and the male subgroup is expressed by $r_1(\bm{x}) = \mathds{1}( x_F = 0)$, where the second term in the product is removed since it is equal to 1.

Decision rules can be leveraged to explore the heterogeneity in the treatment effects via partitioning of the overall sample into subgroups characterized by different HTEs. Decision rules are particularly appealing for this scope, as they guarantee a high level of interpretability while simultaneously exploring potentially complex heterogeneity patterns.

\vspace{1cm}
\begin{minipage}{\textwidth}
\hspace{-1cm}
	\begin{minipage}[]{0.5\textwidth} 
		\centering
		\begin{tikzpicture}[level distance=60pt, sibling distance=10pt, edge from parent path={(\tikzparentnode) -- (\tikzchildnode)}]
			\tikzset{every tree node/.style={align=center}}
			\Tree [.\node[rectangle,draw]{All}; \edge node[auto=right,pos=.6]{$x_F=0$}; \node[rectangle,draw]{Male}; \edge node[auto=left,pos=.6]{$x_F=1$};[.\node[rectangle,draw]{Female}; \edge node[auto=right,pos=.6]{$x_Y=0$}; \node[rectangle,draw]{Old female}; \edge node[auto=left,pos=.6]{$x_Y=1$}; \node[rectangle,draw]{Young female};  ]]
		\end{tikzpicture}
		\label{graph:example_tree}
	\end{minipage}
	\begin{minipage}[]{0.5\textwidth}
		\centering 
		\begin{tabular}{ccc}
			\toprule
			 & Rule & Subgroup \\ [0.05cm]
			\midrule
			$r_1$ & $x_F=0$  & Male\\
			$r_2$ & $x_F=1$ & Female \\
			$r_3$ & $x_F=1$ \& $x_Y=0$ & Old female \\
			$r_4$ & $x_F=1$ \& $x_Y=1$ & Young female\\
			\bottomrule
		\end{tabular}
		\label{tab:rules_example}
	\end{minipage}
\captionof{figure}{An example decision tree (left) and the corresponding decision rules (right).}
\label{fig:example_tree}
\end{minipage}
\vspace{0.3cm}

%\subsection{Pseudo-Outcome Regression with Decision Rules}

Let $\mathcal{R}=\{r_m\}_{m=1}^M$ be a set of decision rules. For each individual $i \in \mathcal{I}$, $\tau_i$ decomposes as
    \begin{equation}
        \tau_i = \bar{\tau} + \sum_{m=1}^M \alpha_m(\mathcal{R}) \cdot r_m(\bm{X}_i) + \nu_i,
    \label{eq:itelindecomposition}
    \end{equation}
where $\bar{\tau}$ is the ATE, $\{\alpha_m\}_{m=1}^M$ are the additive contributions for the activated rules characterizing the heterogeneity, and $\nu_i$ is an unobserved and independent additive noise  (with $\E[\nu] =0 \text{ and }\text{Var}[\nu] = \sigma^2$). In terms of the conditional expectation, equation \ref{eq:itelindecomposition} becomes:
\begin{equation}
\begin{split}
    \tau(\bm{x}) &=\bar{\tau} + \sum_{m=1}^M \alpha_m(\mathcal{R}) \cdot r_m(\bm{x}).
\end{split}
\label{eq:catelindecomposition}
\end{equation}
If all the covariates are discrete, such a decomposition always exists; if not (i.e., in the case of continuous covariates), it can be considered a step-wise approximation. For each rule $r_m$, the coefficient $\alpha_m(\mathcal{R})$, if defined, represents its additive contribution to the CATE, fixing the values of all the other decision rules. In formula:
\begin{equation}
\begin{split}
    \alpha_m(\mathcal{R}) &:= \mathbb{E} [Y_i(1)-Y_i(0)| r_1(\bm{X}_i)=\rho_1, ..., r_m(\bm{X}_i)=1, ..., r_M(\bm{X}_i)=\rho_M] \\
    & - \mathbb{E} [Y_i(1)-Y_i(0)| r_1(\bm{X}_i)=\rho_1, ..., r_m(\bm{X}_i)=0, ..., r_M(\bm{X}_i)=\rho_M]
\end{split} 
\end{equation}
where $\rho_1, ..., \rho_{m-1}, \rho_{m+1}, ..., \rho_{M} \in \{0,1\}$. In particular, it represents the additive contribution to the ATE when all the other rules are set to 0: 
\begin{equation}
    \alpha_m({\mathcal{R}}) := \mathbb{E}\left[Y_i(1)-Y_i(0)| \bm{X}_i \in \left\{\bm{x} \in \mathcal{X}: \left[r_m(\bm{x}) \cdot \prod_{\substack{k=1\\ k\neq m}}^M \left(1-r_k(\bm{x})\right)\right]=1 \right\}\right]- \bar{\tau}
\end{equation}
%In the context of the toy example in Figure \ref{fig:example_tree}, we can imagine $\alpha_m({\mathcal{R}})$ being the additive effect when just one of the four rules is activated and the other are not.

In the rest of the paper, we refer to $\alpha_m(\mathcal{R})$ as the Additive Average Treatment Effect (AATE) of the $m$-th rule, and to simplify notation, we remove its dependency on $\mathcal{R}$. 

Model in equation \eqref{eq:itelindecomposition} can be rewritten in a matrix form as:
\begin{equation}
    \bm{\tau} = \bar{\tau} + \bm{R} \bm{\alpha} + \bm{\nu},
\label{eq:mlindecomposition}
\end{equation}
where $\bm{\tau} \in \mathbb{R}^N$ is the vector of (unobserved) ITE, $\bar{\tau} \in \mathbb{R}$ is the ATE,  $\bm{\alpha} \in \mathbb{R}^M$ is the vector of AATEs, $\bm{\nu} \in \mathbb{R}^N$ is a vector of heteroscedastic and independent additive noise, and $\bm{R} \in \{0,1\}^{N \times M}$ is the decision rules matrix, where each element $R_{i,j}:= r_j(X_i)$ for all $i \in \{1, ..., N\}$ and $j \in \{1, ..., M\}$.

Let us consider $\mathcal{R}=\{r_1,r_2,r_3\}$ from the toy example in Figure \ref{fig:example_tree} (no need to include $r_4$, since it is a linear combination of $r_2$ and $r_3$), and 4 individuals: $\bm{X}_1$ (young male), $\bm{X}_2$ (old male), $\bm{X}_3$ (young female) and $\bm{X}_4$ (old female). The corresponding decision rules matrix is:
\begin{equation*}
  \bm{R}= \bordermatrix{\text{}&r_1 &r_2 & r_3\cr
                \bm{X}_1& 1 & 0 & 0 \cr
                \bm{X}_2& 1 & 0 & 0 \cr
                \bm{X}_3& 0 & 1 & 0 \cr
                \bm{X}_4& 0 & 1 & 1 \cr}.
\end{equation*}

In this framework, interpretable heterogeneous discovery corresponds to finding a set of decision rules $\mathcal{R}$ that satisfy equation \ref{eq:itelindecomposition}, and HTE inference corresponds to making inference on the $\bm{\alpha}$ (AATE) parameters.

%% file: 3_method.tex
\section{Causal Rule Ensemble}
\label{sec:method}

The decomposition of CATE as a linear combination of interpretable decision rules has a potentially large set of solutions. In this paper, our primary objective is to discover a \textit{stable} decomposition of CATE---i.e., robust under reasonable permutations in the data \citep{yu2013stability}. Also, we prioritize \textit{parsimonious} solutions, characterized by a small number of low-complexity decision rules, while simultaneously maintaining the same level of explanatory power \citep{bargagli2022simple}. Once a stable decomposition of CATE has been found, our goal is to estimate and make inference on the HTE in the discovered subgroups.

In this section, we introduce the Causal Rule Ensemble (CRE), a new algorithm for the discovery and inference of heterogeneous causal effects via an interpretable, simple, and stable characterization of CATE as a linear combination of decision rules. Although CRE enforces interpretability in the model via the discovery of subgroups defined as decision rules, the levels of simplicity and stability of the final output of CRE can be directly set by the researcher herself and tailored to the research question at hand.

Assuming the setup described in section \ref{ssec:potentialputcome}, we first divide the observational dataset into two subsamples: a discovery subsample ($\mathcal{I}^{dis}$) and an inference subsample ($\mathcal{I}^{inf}$). In the discovery step, we use $\mathcal{I}^{dis}$ to discover a set of decision rules, $\hat{\mathcal{R}}$, characterizing the HTE. In the inference step, we use $\mathcal{I}^{inf}$ to characterize the CATE via $\hat{\mathcal{R}}$ and provide inference on the estimated effects. The idea of sample splitting is not new in statistics, and the earliest references can be traced back to \cite{stone1974cross} and \cite{cox1975note}. It is now commonly used also in the HTE literature to prevent overfitting, and it is referred to as \textit{honest splitting} \citep{athey2016, lee2021discovering}. 
%In the sensitivity analysis step, we provide confidence intervals to our estimates in case of one or more identifiability assumptions don't hold. 
The following schematic summary illustrates the main steps of the proposed methodology. In the rest of the section, we discuss in detail all the steps of the proposed procedure and its theoretical guarantees. 

\begin{algorithm}
\footnotesize
\caption{Causal Rule Ensemble (CRE)}
\label{alg:cre}
\vspace{0.15cm}
{\bf Inputs:} covariates matrix $\bm{X}$, (binary) treatment vector $\bm{z}$, and observed response vector $\bm{y}$.\\
{\bf Outputs:} (i) a set of interpretable decision rules $\mathcal{\hat{R}}=\{\hat{r}_m\}_{m=1}^M$, 

\hspace{1.46cm} (ii) ATE $\hat{\bar{\tau}}$ and AATEs $\hat{\bm{\alpha}}$ estimates and confidence intervals,

\vspace{0.05cm}
{\bf Procedure:}
\begin{algorithmic}%[1]
    \State $(\bm{X}^{dis},\bm{z}^{dis},\bm{y}^{dis}$), ($\bm{X}^{inf},\bm{z}^{inf},\bm{y}^{inf}) \gets \texttt{HonestSplitting}(\bm{X},\bm{z},\bm{y}) $

    \vspace{0.02cm}
    \noindent
    {\bf i. Discovery}
    \begin{algorithmic}
        \State $\bm{\hat{\tau}}^{dis} \gets \texttt{EstimateIATE}(\bm{X}^{dis},\bm{z}^{dis},\bm{y}^{dis})$ \Comment{e.g. AIPW, CF, CausalBART, S/T/X-Learner} %(subsection \ref{ssec:ite_dis})}
        \State $ \mathcal{\hat{R}'} \gets \texttt{GenerateRules}(\bm{X}^{dis},\bm{\hat{\tau}}^{dis}) $
        \Comment{i.e., tree-ensemble method} %(subsection \ref{ssec:generation})}
        \State $ \mathcal{\hat{R}} \gets \texttt{RulesSelection}(\mathcal{\hat{R}'},\bm{X}^{dis},\bm{\hat{\tau}}^{dis})$ \Comment{e.g., Stability Selection, LASSO} %(subsection \ref{ssec:selection})}
    \end{algorithmic}
    
    % CATE Inference
    \vspace{0.02cm}
    \noindent
    {\bf ii. Inference}
    \begin{algorithmic}
        \State $\bm{\hat{\tau}}^{inf} \gets \texttt{EstimateIATE}(\bm{X}^{inf},\bm{z}^{inf},\bm{y}^{inf})$ \Comment{e.g. AIPW, CF, CausalBART, S/T/X-Learner} %(subsection \ref{ssec:ite_inf})}
        \State $\hat{\bm{\alpha}} \gets \texttt{EstimateAATE}(\mathcal{\hat{R}},\bm{X}^{inf}, \bm{\hat{\tau}}^{inf})$ \Comment{Linear smoothing} %(subsection \ref{ssec:aate})}
    \end{algorithmic}

    %\vspace{0.02cm}
    %\noindent
    %{\bf iii. Sensitivity Analysis}
    %\begin{algorithmic}
     %   \State $\text{C.I.}(\hat{\bm{\alpha}}) \gets \texttt{SensitivityAnalyisis}(\mathcal{\hat{R}},X^{inf}, \bm{z}^{inf}, \bm{y}^{inf})$ \Comment{See Algorithm \ref{alg:sensitivityanalysis}} 
    %\end{algorithmic}
\end{algorithmic}
\end{algorithm}

\subsection{Discovery via Tree-Ensembles}
\label{sec:discovery}

The discovery step itself is divided into a step of \textit{rules generation} and a step of \textit{rules selection}. In the rules generation step, \textit{pseudo-outcomes} are estimated using any causal-machine learning methodology; then, an ensemble of trees algorithm (e.g., random forest) is trained to discover the heterogeneity in the estimated treatment effects via a \textit{fit-the-fit} approach, and a set of candidate decision rules is extracted. In the rules selection step, only a stable subset of the proposed decision rules is selected based on the stability selection algorithm.  

\subsubsection{Rules Generation}
\label{ssec:generation}

For each individual $i\in \mathcal{I}^{dis}$, we estimate the corresponding IATE, $\hat{\tau}^{dis}(\bm{X}_i)$, which---consistently with the literature \cite[see, e.g, ][]{kennedy2020optimal}---we refer to as the \textit{pseudo-outcome}. For simplicity of notation, the pseudo-outcomes are indexed using $\hat{\tau}^{dis}_i$. CRE is model-agnostic with respect to the pseudo-outcome estimators used in this step, and any algorithm can be used, leading to different finite-sample performances (see simulations in section \ref{sec:sym_discovery}). Please refer to the Supplementary Material for a concise overview of the methods used for this task in the paper. 

We detect heterogeneity in the treatment effect by a \textit{fit-the-fit} approach. Once the pseudo-outcome estimates on the discovery sample are obtained, we fit these estimates ($\hat{\bm{\tau}}^{dis}$) from the observed covariates ($\bm{X}^{dis}$) by a tree-ensemble method. Here, we propose to combine both the Random Forest and the Gradient Boosting Machine (GBM) for rules generation, following the parameters' setting described by \cite{friedman2008predictive} and \cite{nalenz2018tree}. Several variants of tree-ensemble methods can be considered, for example, modifying the tree generation's splitting criteria to enforce heterogeneity discovery \cite{powers2018some}. Once the forest is generated, we test, a posteriori, the predictive performance of each terminal node, comparing an error metric for the model with and without that leaf. If performance drops below a certain threshold ($t_{decay}$), the node is discarded (pruned) as not significant for prediction \citep{deng2019interpreting}.

We associate each node in each tree in the resulting forest with the corresponding decision rule obtained by combining the conditions of all its ancestors. Then, we collect the $M''$ most retrieved decision rules as candidate drivers of the HTE ($\hat{\mathcal{R}}''$). Finally, we discard all the extreme or redundant decision rules (see the Supplementary Material for details on this filtering). The filtered set of candidate decision rules $\hat{\mathcal{R}}' \subseteq \hat{\mathcal{R}}''$ is then given in input to the rules selection step. 

By design, the maximal complexity of candidate decision rules can be controlled a priori by the maximal length parameter ($L$) and other stopping criteria in the tree-ensemble method. $L$ can be chosen by the researcher themselves. Larger values of $L$ will lead to deeper trees, which can be useful to explore complex heterogeneity patterns---e.g., the detection of vulnerable individuals in precision medicine. Smaller values of $L$ will produce simpler trees that can be employed to discover HTE at a large granularity---e.g., the detection of vulnerable communities. The filtering criteria are also very useful in practice to preliminary filter irrelevant decision rules and speed up the rules selection step.

\begin{figure}[h!]
	\centering
	\includegraphics[width=\textwidth]{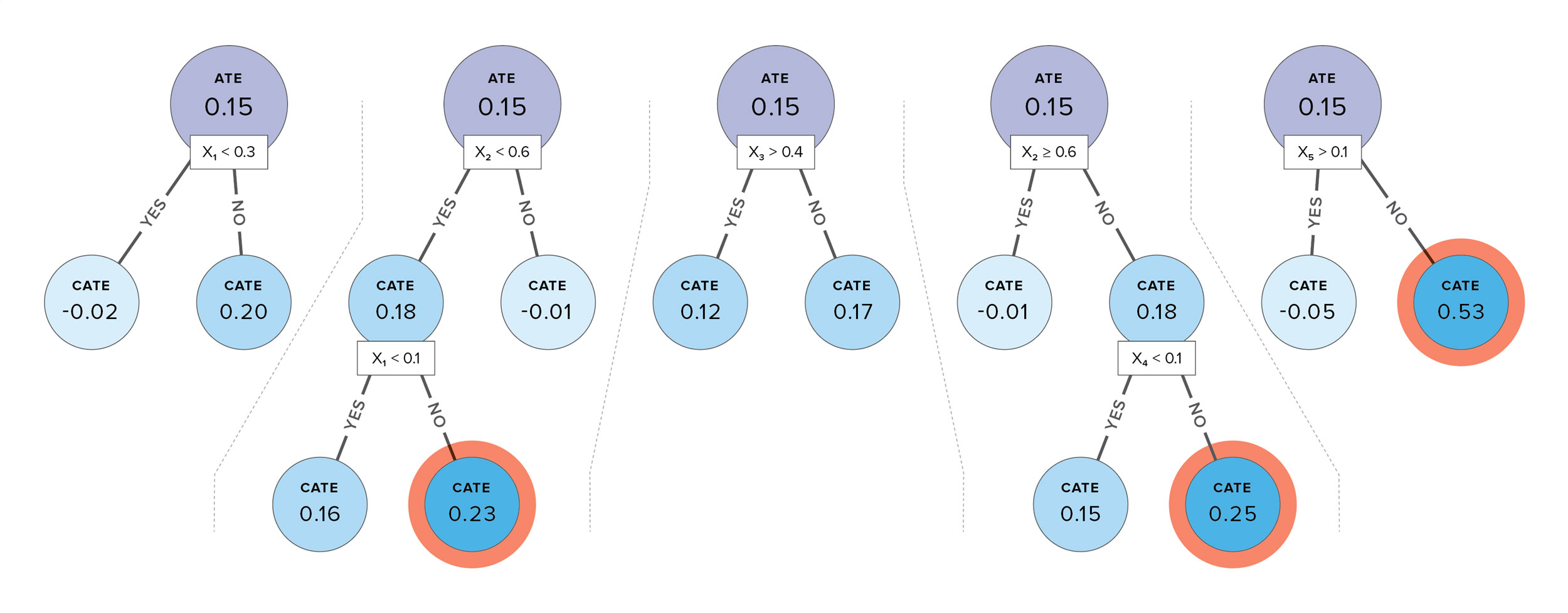} 
	\caption{Visual representation of the rule generation and selection procedure.}
	\label{fig:trees}
\end{figure}

In Figure \ref{fig:trees}, we present an example of a tree ensemble to visualize the procedure described above. The generated forest is made up of $T=5$ trees and 12 total leaves. They correspond to 11 distinct decision rules (``$x_2\geq0.6$" is double). Among these, we discard all the non-significant rules (in light blue) based on the filtering described above. The remaining 8 leaves in dark blue are the candidate decision rules given in input to the rules selection.

\subsubsection{Rules Selection} 
\label{ssec:selection}

The number of candidate decision rules $M'$ extracted by the rules generation procedure grows exponentially with the maximal length and linearly with respect to the number of trees (before filtering). Although the filtering criteria already discard the not-significant rules, we are not ensuring the stability of these decision rules---i.e., given variations in the discovery set, these rules might be replaced with different ones. To enforce stability in discovery, we apply a stability selection regularization procedure to extract only the set of robust and predictive decision rules in terms of heterogeneity characterization. To do so, we rely on the following penalized regression for rule selection 
\begin{equation}
    \underset{\bm{\alpha}}{\mathrm{min}} 
    \left\|\bm{\tau} - (\bar{\tau} + \bm{R}\bm{\alpha}) \right\|_2^2 + \lambda \left\|\bm{\alpha}\right\|_l 
	\label{eqn:modellasso}
\end{equation}
where $R$ is the decision rules matrix, $\lambda \in \Lambda \subseteq \mathbb{R}^+$ is the regularization parameter, and $\left\|\cdot\right\|_l$ is a given norm. We select only the rules whose corresponding AATE estimate is different from zero. The least absolute shrinkage and selection operator (LASSO) estimator \citep{tibshirani1996regression} has been popular and widely used over the past two decades in order to solve the problem in~\eqref{eqn:modellasso} with $l=1$. The usefulness of this estimator among other penalization regression methods is demonstrated in various applications \cite[see, e.g.,][]{su2016identifying, belloni2016inference, chernozhukov2016locally, chernozhukov2017double}. 

In practice, the true IATEs are not observed, and we replace them with the corresponding estimates $\hat{\bm{\tau}}^{dis}$ already computed. We also propose enforcing the discovery of shorter, thus less complex, decision rules by weighting the columns of the rules matrix by the complexity (length) of the corresponding rule:
\begin{equation}
    \tilde{R}_{i,m} = \frac{R_{i,m}}{\text{length}(r_m)} \qquad \forall i \in \mathcal{I}^{dis}, \forall m \in \{1, ..., M'\}.
\end{equation} 

When data is high-dimensional, selecting $\lambda$ can be challenging \citep{hastie2015statistical}. Stability Selection \citep{meinshausen2010stability} provides a procedure to enhance stability in the extraction of the set of decision rules characterizing the HTE while allowing for not explicitly setting the regularization parameter $\lambda$. Here, we propose our rework of this procedure.
Let $\mathcal{D}^{dis}=(\hat{\bm{\tau}}^{dis}, \hat{\mathcal{R}}')$ the pseudo-outcomes and the candidate decision rules estimates just obtained. For each value of $\lambda \in \Lambda$, we bootstrap $B$ different subsample $\mathcal{D}_{(b)}$. For each subsample $\mathcal{D}_{(b)}$ a regularized regression---e.g., the model in \eqref{eqn:modellasso}---is run on $\bm{D}_{(b)}$ to obtain a selection set $\hat{\mathcal{R}}_{(b)}^{\lambda} \subseteq \hat{\mathcal{R}}'$ of decision rules. For each candidate decision rule $r_m \in \hat{\mathcal{R}}'$, let $\pi^{\lambda}_{m}$ its probability of being selected by a certain selection algorithm characterized by $\lambda$:
\begin{equation}
    \pi^{\lambda}_{m} = P \{ r_m \in \hat{\mathcal{R}}^{\lambda} \},
\end{equation}
estimated by:
\begin{equation}
    \hat{\pi}^{\lambda}_{m} =  \frac{1}{B} \sum_{b = 1}^B \mathds{I} \{ r_m \in \hat{\mathcal{R}}_{(b)}^{\lambda} \}. %\approx P \{ r_m \in \hat{\mathcal{R}}^{\lambda} \} .
\end{equation}

Given an estimate of the selection probabilities for each discovered rule and for each value of $\lambda$, we select a stable set of decision rules characterizing heterogeneity in the treatment effect, selecting all rules with selection probability greater than a certain threshold $\pi_{\text{thr}}$ for at least one value of $\lambda$:
\begin{equation}
    \hat{\mathcal{R}} = \{ r_m \: : \max\limits_{\lambda \in \Lambda} \: \hat{\pi}^{\lambda}_m \geq \pi_{\text{thr}} \}
\end{equation}
\cite{meinshausen2010stability} discuss that the empirical results vary little for threshold values $\pi_{\text{thr}} \in (0.6, 0.9)$, and the choice of the set of regularization parameters $\Lambda$ can be explicitly controlled by an upper bound on the (allowed) per-family error rate (PFER). In addition to this, stability selection provides a way to control for false discoveries by bounding the finite sample probability of making a Type I error---i.e., the probability of discovering a false positive decision rule. %Also, \citet{bodinier2021automated} has recently proposed an automated selection of these parameters coming from the maximization of a stability measure.

Using the example above, illustrated in Figure \ref{fig:trees}, among the eight candidate decision rules associated with the terminal nodes in dark blue, the three terminal nodes highlighted in orange represent the selected decision rules by the stability selection procedure.

\subsection{Inference}
\label{sec:estimation}

Once a set of (robust) decision rules $\hat{\mathcal{R}}$ is estimated from the discovery subsample, we estimate the coefficient of the corresponding pseudo-outcome regression on the inference subsample $\mathcal{I}^{inf}$ through a two-stage estimation procedure. 

%\subsubsection{Pseudo-Outcomes Estimation for Inference}\label{ssec:ite_inf}

%\subsubsection{AATE Estimation via Pseudo-Outcome Regression} \label{ssec:aate}

In this section, we present a general two-stage regression approach that regresses estimated \textit{pseudo-outcomes} onto a vector of interpretable decision rules. Prominent examples of previous research that encompass general forms of pseudo-outcomes can be found in \cite{ai2003efficient}, \cite{foster2019orthogonal} and \cite{kennedy2020optimal}. In the context of this paper, this two-stage regression approach is used to decompose the HTE into the linear combination of multiple decision rules.

First, for each individual $i\in \mathcal{I}^{inf}$, we estimate the corresponding IATE, $\hat{\tau}^{inf}(\bm{X}_i)$, which we refer to as the \textit{pseudo-outcome}. For simplicity of notation, the pseudo-outcomes are indexed using $\hat{\tau}^{inf}_i$. As in the discovery phase, CRE is agnostic with respect to the estimators used, and any algorithm can be combined with it, leading to different convergence properties. Note that it is not required to use the same estimator used in the discovery step, and certain methods could be preferred for one task or the other. %For a brief overview of the pseudo-outcome estimators compared in the simulations, see the Supplementary Material.

Based on the model in equation \eqref{eq:catelindecomposition}, we estimate the AATE via Ordinary Least Squares:
\begin{equation}
	\hat{\bm{\alpha}} = (\bm{R}^T \bm{R})^{-1} \bm{R}^T (\hat{\bm{\tau}}^{inf} - \hat{\bar{\tau}}^{inf}),
	\label{eq:OLSestimator}
\end{equation}
discarding the not-significant or redundant decision rules---i.e., covariate t-test with $p$-value $>$ 0.05.
Under a few additional conditions, we prove the consistency and asymptotic normality of the estimator $\hat{\bm{\alpha}}$.

%\subsubsection{Statistical Properties}

\begin{proposition}
\label{prop:consistency} 
Assume (1) and (2). Let $\hat{\bm{\tau}}$ be a consistent estimator for $\bm{\tau}$. Let $\E(\bm{R}^T \bm{R}) = Q$ a positive definite matrix, then the CRE estimator defined in equation \eqref{eq:OLSestimator} is a consistent estimator for $\bm{\alpha}$ (AATE).

\vspace{0.5cm}
\noindent [See proof in the Supplementary Material]
\end{proposition} 
	
\begin{proposition}
\label{prop:alpha_asnormality}
Under Proposition \eqref{prop:consistency} assumptions, and let $\E(\nu_i^4) < \infty$ and $\E(\nu_i^2 \bm{R}_i^T \bm{R}_i) = \Omega \succ 0$ (positive definite) for all $i \in \mathcal{I}^{inf}$ (additional regularity assumptions),
then:
\begin{equation}
    \sqrt{N} (\hat{\bm{\alpha}} - \bm{\alpha})  \overset{d}{\to}  \mathcal{N}(\bm{0}, \bm{V}) \:\:\:  \text{as} \:\:\: N \to \infty
\end{equation}
where $\bm{R}_i$ represents the $i$-th row of the rules matrix $\bm{R}$ and $\bm{V} = \bm{Q}^{-1} \bm{\Omega} \bm{Q}^{-1}$.

\vspace{0.5cm}
\noindent [See proof in the Supplementary Material]
\end{proposition}

A variance-covariance matrix estimator $\hat{\bm{V}} = \hat{\bm{Q}}^{-1} \hat{\bm{\Omega}} \hat{\bm{Q}}^{-1}$ can be obtained by the sandwich formula where:
\begin{align}
    \hat{\bm{Q}} &= \frac{\bm{R}^T \bm{R}}{N}, \\
    \hat{\bm{\Omega}} &= \frac{\bm{\nu}^T \bm{R} \bm{R}^T \bm{\nu}}{N}, \\
    \hat{\bm{\nu}} &= \hat{\bm{\tau}}^{inf} - (\hat{\bar{\tau}}^{inf} + \bm{R} \bm{\alpha}),
\end{align}
This can be used to provide confidence intervals for the AATE estimates. This estimator is robust and often referred to as White's estimator \citep{white1980heteroskedasticity}. There are other approaches to obtain a heteroskedasticity-consistent covariance matrix as discussed in \cite{long2000using}. For small samples, Efron's estimator \citep{efron1982jackknife}, known as the HC3 estimator, can be considered alternatively. Also, if the variance $\sigma_i^2$ is known from the large sample properties of existing methods to obtain $\hat{\tau}_i^{inf}$, then feasible generalized least squares estimators \citep{lewis2005estimating} can be considered. 
In order to also take into account the uncertainty propagation from the pseudo-outcome estimation, we also considered uncertainty quantification by bootstrapping, sampling $B$ subsamples from $\mathcal{I}^{inf}$ and averaging the obtained coefficients.

We note that the proposed pseudo-outcome regression is similar in spirit to the one proposed by \citet{kennedy2020optimal} where equation \eqref{eq:itelindecomposition} can be interpreted as a linear smoother for which the notion of stability in Definition 1 in \citet{kennedy2020optimal} holds. However, there are notable distinctions from the approach proposed by \cite{kennedy2020optimal} that should be highlighted. Firstly, our target estimands differ as we employ a collection of interpretable decision rules rather than relying solely on covariates. Secondly, we maintain a model-agnostic perspective concerning the estimation of pseudo-outcomes, whereas Kennedy's approach primarily focuses on employing the augmented inverse propensity weighting estimator \citep{robins1994estimation}. %Here, we maintain a more flexible stance by allowing for various choices of models for estimating the pseudo-outcomes.

\subsection{Sensitivity Analysis to Unmeasured Confounding}\label{subsec:confounding}
The estimator $\hat{\bm{\alpha}}$ succinctly encapsulates the heterogeneity of the treatment effects. Its validity and consistency are based on the assumption~\ref{ass:ignorability} and the correct specification of the propensity score. However, in practical scenarios, it remains uncertain whether the chosen set $\bm{X}_i$ adequately satisfies Assumption~\ref{ass:ignorability}. Our proposed approach can be supplemented by sensitivity analysis evaluating the robustness of the causal conclusion to the impact of potential unmeasured bias. More specifically, it is straightforward to adopt the marginal sensitivity model that was introduced by \cite{tan2006distributional} and \cite{zhao2019sensitivity}. This model focuses on two aspects: (i) the true propensity probability $e_0(\bm{x}, y; z) = P_0(Z=1|\bm{X}=\bm{x}, Y(z) = y)$ and (ii) within a parametric model framework $e_{\gamma}(\bm{x})$, the best approximation/projection of $e_0(\bm{x})$, denoted as $e_{\gamma_0}(\bm{x})$. Our sensitivity analysis model assumes that the odds ratio between $e_0(\bm{x}, y; z)$ and $e_{\gamma_0}(\bm{x})$ lies within the range $[1/\Lambda, \Lambda]$ for $z \in \{0,1\}$ and some $\Lambda \geq 1$. A condition where $\Lambda = 1$ indicates $e_0(\bm{x}, y; z) = e_{0}(\bm{x}) = e_{\gamma_0}(\bm{x})$ for all $z$, suggesting that there is no deviation from the initial assumptions and that the propensity model is correctly specified. This approach allows for an assessment of the deviations simultaneously. Further elaboration on the sensitivity analysis model is provided in the Supplementary Material section~\ref{app:sensi}. 

%% file: 4_simulations.tex
\section{Simulations}
\label{sec:simulations}

To assess the relative performance of CRE, we carried out two simulation studies aimed at evaluating the performance of CRE in its ability to (i) correctly discover the HTE and (ii) precisely estimate the CATE. In the first simulation study, we evaluated the performance in heterogeneity characterization, retrieving the correct effect modifiers and decision rules. We compare different variants of CRE using different pseudo-outcome estimators and evaluate them with different magnitudes of the causal effect. In the second simulation study, we benchmark the accuracy of the CRE estimation. We compare the CRE estimator with state-of-the-art methodologies for the HTE estimation. Empirical verification of the consistency of AATE estimation (Proposition \ref{prop:consistency}) is included in the Supplementary Material.

We consider several data-generating processes, varying the confounding mechanism, sample size, and the number and complexity of the rules, and comparable results are obtained. In this section, we report the main results of both analyses; for a complete overview of the results with the full battery of data-generating processes, see the Supplementary Material.

\subsection{Discovery}
\label{sec:sym_discovery}

Let $\mathcal{I}$ be a sample of $N=2,000$ individuals. For each individual $i \in \mathcal{I}$, let us define: $X_i^1,..X_i^{p} \overset{\mathrm{iid}}{\sim} Bernoulli(0.5)$, and $Z_i\sim Bernoulli(\pi_i)$ with $ \pi_i=\frac{1}{1+e^{+1-X_i^1+X_i^2-X_i^3}}$ where $\bm{X}_i$ is the vector of the $p=10$ observed (binary) covariates of the individual $i$, and $Z_i$ represents its assigned (binary) treatment. Let's further define the potential outcomes: 
$$Y_i(0) \sim \mathcal{N}(\mu_i^0,1) \quad \text{with} \quad \mu_i^0 = f(\bm{X}_i) +k \cdot \mathds{1}_{\{x_1=1;x_2=0\}}(\bm{X}_i),$$ $$Y_i(1) \sim \mathcal{N}(\mu_i^1,1) \quad \text{with} \quad \mu_i^1 = f(\bm{X}_i) +k \cdot \mathds{1}_{\{x_5=1;x_6=0\}}(\bm{X}_i)$$ where $k \in \mathbb{R}$ represents the magnitude of the causal effect and $f$ is a linear form (by default $f(\bm{X}_i)=X_i^1+X_i^3+X_i^4$). The linearity of $f$ is relaxed in the Supplementary Material with no notable changes in the algorithm's performance. 
It follows that the (unobserved) treatment effect for individual $i$ is equal to $\tau_i = Y_i(1)-Y_i(0) =  - k \cdot \mathds{1}_{\{x_1=1;x_2=0\}}(\bm{X}_i) + k \cdot \mathds{1}_{\{x_5=1;x_6=0\}}(\bm{X}_i) + \nu_i$
where $\nu_i \sim \mathcal{N}(0,2)$ is an additive zero-mean noise, with $M=2$ decision rules, $\bar{\tau}=0$, $r_1(\bm{x})=\mathds{1}_{\{x_1=1;x_2=0\}}(\bm{x})$, $r_2(\bm{x})=\mathds{1}_{\{x_5=1;x_6=0\}}(\bm{x})$,  $\alpha_1= \alpha_2 = k$, and $\tau(\bm{x}) =  - k \cdot \mathds{1}_{\{x_1=1;x_2=0\}}(\bm{x}) + k \cdot \mathds{1}_{\{x_5=1;x_6=0\}}(\bm{x}) = \sum_{m=1}^M \alpha_m \cdot r_m(\bm{x})$. This case is of particular interest because it corresponds to a scenario in which the simple investigation of the ATE would lead to a null effect despite a notable heterogeneity in the treatment effects.

We measure the CRE's capability in retrieving both the true effect modifiers (i.e., $x_1, x_2, x_5, x_6$) and the exact decision rules (i.e., $r_1, r_2$) varying the magnitude of the causal effect (i.e., varying $k$). 
Let $\mathcal{A}$ be the set of true effect modifiers (or decision rules) and $\hat{\mathcal{A}}$ the set discovered by CRE. We first define the number of elements correctly retrieved $TP  =|\hat{\mathcal{A}}\cap\mathcal{A}|$, the number of elements incorrectly retrieved $FP  =|\hat{\mathcal{A}}-\mathcal{A}|$, and the number of true elements not retrieved $FN =|\mathcal{A}-\hat{\mathcal{A}}|$. Thus, we can then define the $Recall  =\frac{TP}{TP+FN}$, $Precision =\frac{TP}{TP+FP}$, and \textit{F1-score} $= 2 \cdot \frac{Recall \cdot Precision}{Recall + Precision}$; where $Recall$ is the ratio of true elements properly retrieved (\textit{quantitative} performance), the $Precision$ is the ratio of correct elements retrieved (\textit{qualitative} performance), and the \textit{F1-score} combines these two measures into a harmonic mean. 

We consider six variants of CRE with six distinct pseudo-outcome estimators: Augmented Inverse Probability Weighting (AIPW), Bayesian Causal Forest (BCF), Causal Bayesian Additive Regression Trees (Causal BART), S-Learner, T-Learner, and X-Learner (see the Supplementary Material for a complete overview of these methods). For each variant of CRE and the size of the causal effect $k$, we compare the mean $Recall$, $Precision$, and \textit{F1-score} and their corresponding 95\% confidence intervals over 250 Monte Carlo experiments. 

\begin{figure}[h!]
	\centering
	\includegraphics[width=\textwidth]{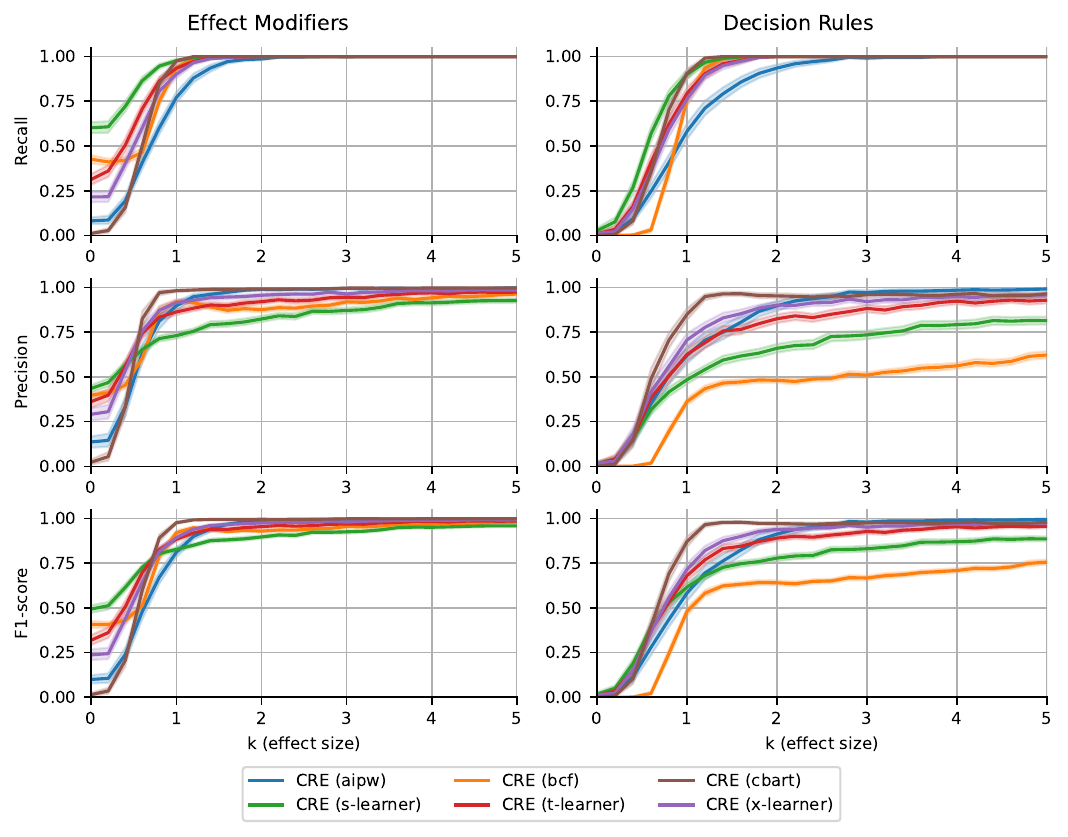} 
	\caption{Simulation study for heterogeneity discovery results with 2 rules, linear confounders, and 2,000 observations. Mean $Precision$, $Recall$ and \textit{F1-score} (lines) with the corresponding 95\% confidence intervals (bands) over 250 Monte Carlo experiments are reported for each method and causal effect size $k$. %For each CRE variant, the heterogeneity characterization discovery converges (with respect to effect size) to the true heterogeneity characterization.
 }
	\label{fig:discovery_main}
\end{figure}

In Table \ref{tab:parameters} in the Supplementary Material, we summarize the parameters of the Causal Rule Ensemble method and the hyperparameters used for this analysis.
The results are reported in Figure \ref{fig:discovery_main}. 
As expected, both effect modifiers retrieval and decision rules retrieval metrics increase monotonically with respect to the causal effect size $k$. All the method variants perform similarly for the retrieval of effect modifiers, and all variants reach almost perfect discovery by $k=3$. Decision rule discovery is more challenging due to the larger hypothesis space. Indeed, in the case of binary covariates (current setting), the number of possible decision rules is equal to $\sum_{l=1}^{L} \binom{p}{l}\cdot 2^l$ growing exponentially with the maximum rules' length $L$, and even more in the setting of discrete or continuous covariates, depending on the discretization criteria in the rules generation. All the method variants retrieve all the true decision rules (Recall$=1$) by $k=3$, which suggests that no further candidate rules should be proposed in the Rules Generation step. However, a few variants (i.e., BCF and S-Learner) also keep retrieving additional non-informative rules (Precision $\ll 1$) even for $k>3$. This drawback can be addressed by fine-tuning the strength in the Rules Selection step (e.g., increasing cutoff $\pi_{\text{thr}}$ in the Stability Selection), which was kept constant for all the methods for a fair comparison. Causal Rule Ensemble based on Causal BART, AIPW, T-Learner, and X-Learner for pseudo-outcome estimation leads to a more stable and precise rules discovery. In agreement with the literature on the excellent performance of Causal BART in causal inference tasks with small effect magnitude \citep{dorie2019automated, hahn2019atlantic, wendling2018comparing}, CRE based on Causal BART still works better than any other variants in such regime.
%Consistent results are obtained when varying: (i) the sample size (1,000, 2,000, 5,000); (ii) the number of the decision rules (2, 4); (iii) the complexity of the decision rules (1, 2, 3); (iv) the type of confounding (none, linear, non-linear). A comprehensive analysis of these additional simulations is reported in the Supplementary Material.

\subsection{Estimation}
\label{sec:sym_estimation}

The simulation study on the HTE estimation follows the same data-generating process described in section \ref{sec:sym_discovery}. For each method, we evaluate the mean and standard deviation Root-Mean-Square Error (RMSE) and Bias on pseudo-outcome estimation over 250 Monte Carlo experiments per method. We consider the same six variants of CRE with the following pseudo-outcome estimators: Bayesian Causal Forest (BCF), Causal Bayesian Additive Regression Trees (Casual BART), S-Learner, T-Learner, and X-Learner; and we compare them with the same ``standalone'' pseudo-outcome estimators, which are often recognized among the best-performing algorithms for heterogeneous treatment effect estimation. We report the results in Table \ref{tab:estimation_main}.

\begin{table}[h!]
\centering
\begin{tabular}{crrrr}
\hline
 & \multicolumn{2}{c}{RMSE} & \multicolumn{2}{c}{Bias} \\
Method & \multicolumn{1}{c}{$\mu$} & \multicolumn{1}{c}{$\sigma$} & \multicolumn{1}{c}{$\mu$} & \multicolumn{1}{c}{$\sigma$} \\ \midrule \midrule
CRE (AIPW) & \textbf{0.1336} & 0.0603 & 0.0016 & 0.0891 \\
\rowcolor{gray!10} CRE (BCF) & 0.1482 & 0.0558 & 0.0047 & 0.0795 \\
CRE (S-Learner) & 0.1494 & 0.0589 & 0.0017 & 0.0860 \\
\rowcolor{gray!10} CRE (T-Learner) & 0.1495 & 0.0649 & 0.0011 & 0.0937 \\
CRE (X-Learner) & 0.1466 & 0.0659 & 0.0010 & 0.0937 \\
\rowcolor{gray!10} CRE (Causal BART) & 0.1398 & 0.0625 & \textbf{0.0009} & 0.0816 \\
\hline
AIPW & 2.0807 & 0.1919 & 0.0032 & 0.0562 \\
\rowcolor{gray!10} BCF & 0.1339 & 0.0373 & 0.0042 & 0.0522 \\
S-Learner & 0.4837 & 0.0334 & 0.0020 & 0.0532 \\
\rowcolor{gray!10} T-Learner & 0.8065 & 0.0373 & 0.0035 & 0.0573 \\
X-Learner & 1.1878 & 0.0291 & 0.0035 & 0.0573 \\
\rowcolor{gray!10} Causal BART & 0.9925 & 0.0163 & 0.0020 & 0.0520 \\ \hline
\end{tabular}
\caption{Simulation study for (heterogeneous) treatment effect estimation, with $M=2$ rules, linear confounder, 2,000 individuals, and under CATE linear decomposition assumption. For all the methods, the mean ($\mu$) and standard deviation ($\sigma$) treatment effect root mean squared error (RMSE) and bias (Bias) over 250 Monte Carlo experiments are reported.}
\label{tab:estimation_main}
\end{table}

Overall, CRE outperforms the corresponding `standalone' pseudo-outcome estimators. In particular, CRE (AIPW), CRE (S-Learner), CRE (T-Learner), CRE (X-Learner), and CRE (Causal BART) outperform the corresponding AIPW, S-Learner, T-Learner, X-Learner, Causal BART estimators for pseudo-outcome estimation, and CRE (BCF) and BCF lead to comparable performances. Among the `standalone' pseudo-outcome estimators, BCF is the method with the largest Bias ($>0.04$). Our hypothesis is that its corresponding errors in pseudo-outcome estimation in the CRE discovery step lead to incorrect heterogeneity characterization, propagating the error in the inference step.

%Consistent results are obtained varying: (i) the sample size (1,000, 2,000, 5,000); (ii) the number of the decision rules (2, 4); (iii) the decision rules' complexity (1, 2, 3); (iv) the type of confounding (none, linear, non-linear). A comprehensive analysis of these additional simulations is reported in the Supplementary Material.

%% file: 5_application.tex
\section{Heterogeneous Effects of Air Pollution Exposure on Mortality}
\label{sec:application}

The literature indicates that long-term exposure to lower levels of fine particulate matter, also known as PM$_{2.5}$, is associated with a significant decrease in mortality \cite[see][for a review]{epa2022}. Although there is a very extensive research on the estimation of population average effects of long-term PM$_{2.5}$ exposure on mortality, very limited research has provided data-driven discovery of the set of effect modifiers that would lead to heterogeneous causal effects. However, it is essential to investigate how the causal effect may differ between different groups of individuals in health studies to answer the call of EPA \citep{epa2022a, epa2022b} to discover environmental inequalities and to pave the way for the development of more effective health policies.

In this context, our focus is on identifying vulnerability or resilience in the causal effects with respect to the population-average causal effects of exposure to air pollution on mortality. In particular, we examine the heterogeneous effects of long-term exposure to high levels of PM$_{2.5}$ among 65-year-old people and older who were enrolled in Medicare in the years 2010-2016. Using CRE, we show how our approach can identify different groups, estimate the heterogeneity in the effects of long-term PM$_{2.5}$ exposure on mortality, and identify the characteristics that distinguish the different heterogeneous subgroups.

\subsection{Data}

We collected records from 35,331,290 Medicare beneficiaries across the contiguous U.S. For each beneficiary, we have information on age, sex, race (specifically categorized as non-Hispanic White, Black/African American, Asian/Pacific Islander, American Indian/Alaska native, and other race), eligibility for Medicaid (which is a proxy of low social-economic status), ZIP code of residence, and an indicator of mortality in the five follow-up years (2012-2016). To capture the long-term effects of exposure to air pollution, we averaged all daily estimates of ambient $\text{PM}_{2.5}$ across 2010 and 2011. $\text{PM}_{2.5}$ exposure estimates were obtained from a well-validated model that estimates daily $\text{PM}_{2.5}$ exposure levels in each 1 km x 1 km grid over the contiguous U.S. \citep{di2019ensemble}. To obtain exposure estimates at the ZIP-code level, estimates from all grids with centroids that fell within a given ZIP code were averaged. %Figure \ref{fig:map_pm25} depicts the average levels PM$_{2.5}$ for the biennium 2010-2011 across the contiguous U.S.

\begin{figure}[h!]
	\centering
	\includegraphics[width=\textwidth]{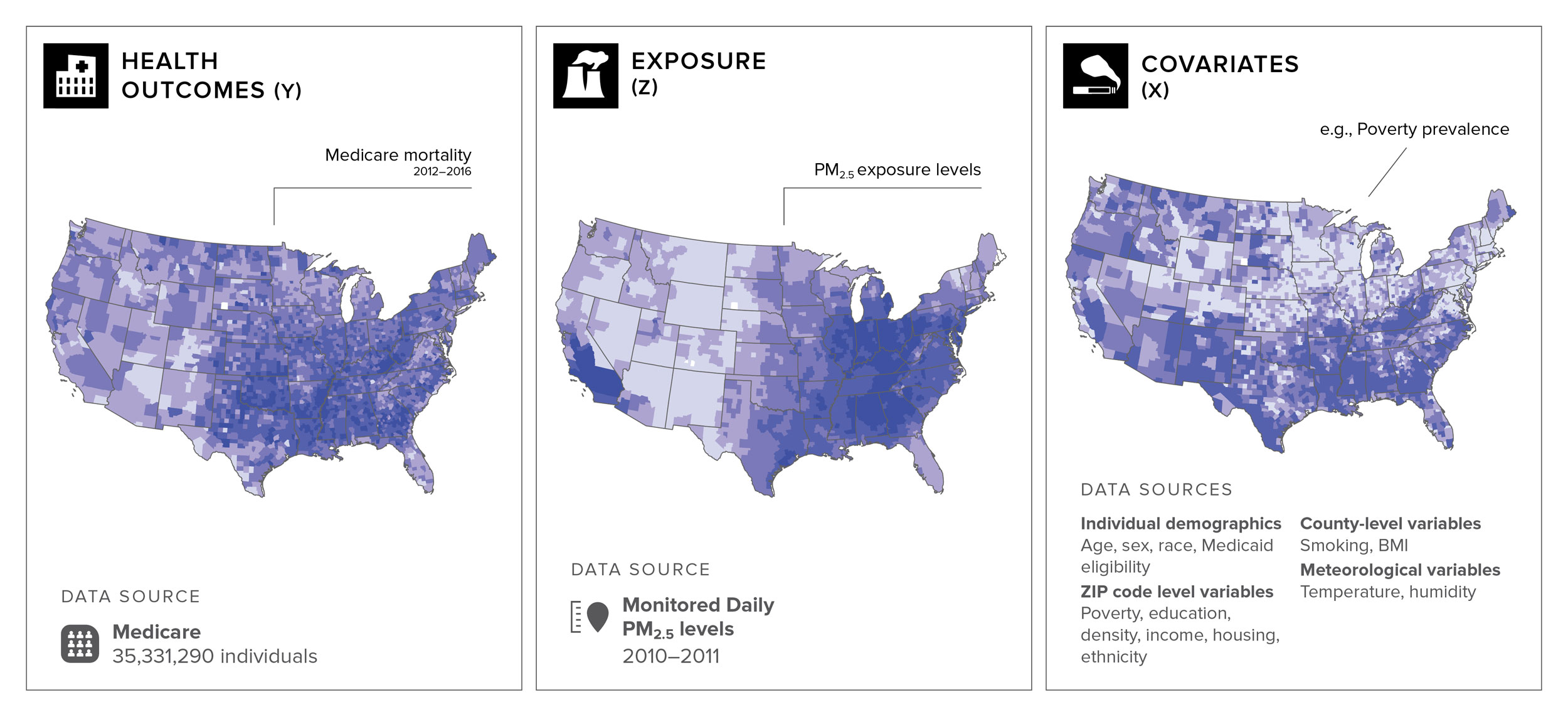} 
	\caption{Summary and visualization of the resulting data set.}
	\label{fig:medicare_dataset}
\end{figure}

Furthermore, to further control for possible sources of confounding, we integrated neighborhood-level variables collected by the U.S. Census, American Community Survey, and the Behavioral Risk Factor Surveillance System. We included several sociodemographic variables, which were captured at the ZIP-code level: average household income, average home value, the proportion of residents in poverty, the proportion of residents without a high school diploma, the population density, the proportion of residents who own their houses, the proportion of the population that is Black, and the proportion of the population that is Hispanic. In addition, we considered average maximum daily temperatures and relative humidity during summer (June to September) and winter (December to February), which were also recorded at the ZIP code level. At the county level, we considered average body mass index and smoking rate, which we then mapped to the appropriate ZIP codes. A summary and visualization of the resulting dataset are reported in Figure \ref{fig:medicare_dataset}.

\subsection{Study Design}
\label{sec:study_design}

We define the treatment variable as $Z=1$ if the average PM$_{2.5}$ in 2010 and 2011 is above the threshold of $12 \mu g/m^3$ and $Z=0$ otherwise. The choice of $12 \mu g/m^3$ as a threshold is in accordance with the National Ambient Air Quality Standard (NAAQS) established by the Environmental Protection Agency (EPA) during the study period. Thus, this binary exposure is extremely relevant from a policy perspective and a similar policy-relevant dichotomiation of PM$_{2.5}$ has been previously explored in the literature \cite[see, e.g.,][]{lee2021discovering}. %The highest proportion of individuals were exposed to PM$_{2.5}$ above $12 \mu g/m^3$ in the Midwest, where 19\% of individuals were exposed. In the West, South, and Northeast, 8\%, 7\%, and 6\% of the individuals were exposed to PM$_{2.5}$ above this threshold, respectively.

All covariates at the individual level (except for age) are already binary (or categorical and easily binarized), and we keep them as such. To further enforce an explicable characterization of HTE, we binarize all the other covariates---i.e., age, ZIP code level variables, and county level variables--- into low and high values using the median as a threshold. For each individual, the observed factual outcome $Y$ is equal to 1 if the person died in the five following years (2012-2016) and 0 otherwise.

We investigate the heterogeneity in the effects of air pollution on mortality separately for the four geographical regions defined by the U.S. Census Bureau: Northeast, Midwest, West, and South (see Figure \ref{fig:regions} in the Supplementary Material for a map of the four regions). It is crucial to investigate the effects of air pollution on health across the different geographical regions of the U.S. for several reasons. Firstly, the U.S. is a vast country with a diverse climate and environmental conditions, leading to substantial differences in air quality and exposure to air pollution across different regions. As highlighted by \cite{baxter2013examining}, it is utterly important to assess the differential risks of air pollution on mortality at a regional level, as they could be suggestive of heterogeneous health responses driven by variations in the PM$_{2.5}$ composition and the concentration of gaseous pollutants. Secondly, people living in different regions may have different susceptibilities to the health effects of air pollution due to various factors such as genetics, lifestyle, and pre-existing health conditions \citep{kloog2013long, zanobetti2009fine}. Therefore, understanding the heterogeneity in the health effects of air pollution across different regions can help identify vulnerable populations and design targeted interventions to mitigate the adverse health effects. Thirdly, \cite{dedoussi2020premature} found that 41 to 53 percent of air-quality-related premature mortality resulting from a state's emissions occurs outside that state. Hence, regional-level analyses---factoring in some of the potential out-of-state sources of emission---directly map into region-wide policies that may be more effective in reducing the mortality burdens from exposure to air pollution. All this considered, investigating the effects of air pollution on health across different regions of the U.S. is essential for identifying the specific risks associated with exposure to pollutants, understanding the heterogeneity in the health effects across different populations, and informing public health policies and interventions.

Verification of no spatial auto-correlation within regions is verified by Moran's test in the Supplementary Material. The list of CRE hyper-parameters used in these analyses is reported in the Supplementary Material.

\subsection{Results}
\label{sec:results}

Consistently with the literature, we found that being exposed to higher levels of air pollution compared to the previous NAAQS of $12 \mu g/m^3$ leads to an increase in mortality in each of the four regions of the contiguous U.S. considered. The greatest increase in the average effect of treatment was found in the Northeast, where individuals exposed, in the biennium 2010-2011, to levels of PM$_{2.5}$ higher than the NAAQS were found to have a 16.2\% (95\% CI: 16.1\% to 16.3\%) higher risk of dying in the five following years, compared to if they were exposed to levels lower than the NAAQS. We found 14.9\% (14.8\% to 15.1\%), 7.1\% (6.9\% to 7.2\%), and 2.3\% (2.2\% to 2.4\%) increases in mortality in the West, Midwest, and South, respectively.

\begin{figure}[h!]
	\hspace{-0.5cm}
	\includegraphics[width=\textwidth]{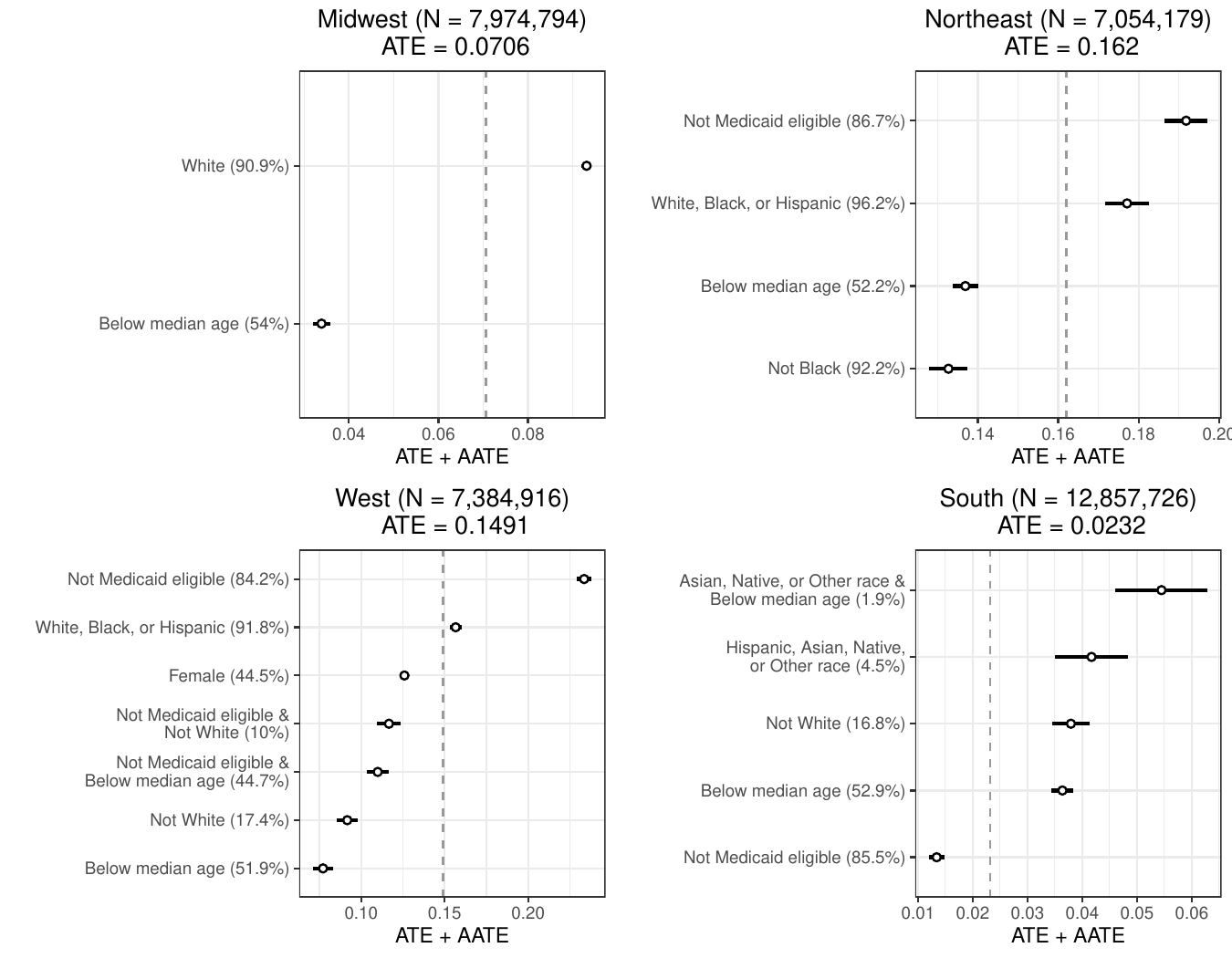} 
	\caption{Results obtained from CRE for the discovery and estimation of HTE of air pollution exposure on mortality. For each U.S. Census geographical region, at the top of the plots, we report the number of people living in that region ($N$) and the Average Treatment Effect (ATE). The dotted line represents the value of the ATE. Points report values of Additive Average Treatment Effect (AATE) for the discovered decision rules (with the corresponding 95\% confidence interval). The results are presented in terms of ATE + AATE to facilitate the interpretation of the coefficients. Additionally, we report the percentage of people who meet the criteria for each decision rule.} 
	\label{fig:medicare_result}
\end{figure}

%For each of the four regions, we found positive and negative AATEs. When the AATE is positive, it indicates that, if that decision rule is activated, the estimated treatment effect is greater than if that decision rule was not activated (while keeping all other factors constant). When an AATE is negative, vice versa. For instance, an AATE of 0.05 for Medicaid eligibility in the Midwest means that Medicaid-eligible people in the Midwest have an additive increase of 5\% in the risk of dying (with respect to the ATE) in the five following years than those not eligible, ceteris paribus. 

Using CRE, in addition to the ATE, we also discovered notable heterogeneity with respect to the AATE in each of the four regions. We used only individual-level variables to define the different subgroups. Figure \ref{fig:medicare_result} shows the results of our analyses. The subgroups whose ATE + AATE coefficients---reported on the horizontal axis of the four plots---are on the right side of the ATE (the dotted line in the plots) depict an increased vulnerability with respect to the average effect. In contrast, the subgroups with ATE + AATE coefficients on the left side of the ATE have a lower vulnerability with respect to the ATE. It is critical to note that for all regions we find a systematic detrimental impact of exposure to higher levels of air pollution on health also at a subgroup level. In fact, the coefficients of the ATE + AATE are never below zero. This indicates that for each of the discovered subgroups there is an increase in mortality following exposure to levels of PM$_{2.5}$ higher than NAAQS.

Let us start from the subgroups that depict higher-than-the-average vulnerability---i.e., ATE + AATE on the right side of the dotted ATE line. In all four regions, we see relationships between demographic and socioeconomic characteristics and vulnerability to air pollution, although these relationships are complex and differ by region. In the Midwest, we find increased susceptibility among non-Hispanic White individuals; in the Northeast and South, we see this among non-Hispanic White, Black and Hispanic beneficiaries, while in the South, all non-White individuals were determined to be higher risk. Being ineligible for Medicaid was also found to be a driver of increased vulnerability in the Northeast and West. Previous studies that have explored vulnerability to air pollution in the Medicare population have identified increased susceptibility in marginalized groups \cite[see, e.g.,][]{di2017air, josey2023air}. Surprisingly, vulnerability in non-marginalized groups has also been documented in the literature \citep{liu2021recent}. % This finds a possible explanation in survival bias \citep[see, e.g.,][]{mayeda2018does, shaw2021evaluation}. Survival bias occurs in studies with older cohorts, such as Medicare, in which some of the most vulnerable individuals, but particularly those in marginalized groups, die before becoming eligible to join the cohort at age 65. This means that those in marginalized groups who do ultimately enter the cohort may be especially resilient and could have a lower mortality risk, with respect to the average risk, even when exposed to higher levels of pollutants \citep{liu2021recent}.
 
In juxtaposition, the groups that show a lower-than-the-average vulnerability---i.e., ATE + AATE on the left side of the dotted ATE line---are mainly composed of a population with younger individuals (in particular in the Midwest, West, and Northeast). Sex was discovered only in the West, where female beneficiaries were found to be less vulnerable than male beneficiaries. Both these results are also in line with what was previously found in the literature \cite[see, e.g.,][]{di2017air, lee2021discovering}. Interestingly, we find that being ineligible for Medicaid lowers the vulnerability to PM$_{2.5}$ in the South and in the West (in combination with being non-White and being below median age). Results for ineligibility for Medicaid should be further investigated as Medicaid eligibility guidelines vary across the U.S.

In conclusion, this application demonstrates the ability of CRE to retrieve non-trivial, yet interpretable, characterization of the heterogeneous subgroups. Furthermore, many of the discovered subgroups had already been identified as potential drivers of vulnerability and resilience in the previous literature. CRE is able to confirm these findings in a data-driven way, which were originally obtained using traditional statistical techniques.

%% file: 6_conclusion.tex
\section{Conclusion}
\label{sec:conclusion}

In this paper, we introduce a new method for interpretable discovery and estimation of heterogeneous treatment effects. The proposed CRE methodology accommodates the well-known shortcomings in the flexibility of (individual) causal trees and interpretability of ensembles of causal trees (i.e., causal forest) relying on the linear decomposition of the treatment effect in terms of decision rules. The heterogeneity discovery is simple and stable to sample-to-sample variations, and under the identifying assumptions, CRE leads to consistent estimates.

The decision rules characterizing heterogeneity are discovered by a \textit{fit-the-fit} procedure, relying on a preliminary estimate of the individual average treatment effect (pseudo-outcome) by any existing treatment effect estimator. Similarly, the final linear model relies on an analogous preliminary individual average treatment effect (pseudo-outcome) estimation. Different properties characterize different estimators, and the performance of CRE varies with respect to them. If an estimator properly estimates the heterogeneity in the treatment effect, the CRE method discovers the underlying structure of the treatment effect with higher probability, representing this structure in an easy-to-interpret form. 

The maximal number and complexity of the rules can be set by researchers or practitioners. Indeed, a few simple (i.e., not lengthy) rules are utterly important for public policy implications, where heterogeneity discovery, potentially informing policy guidelines, needs to be as simple and as general as possible. However, when it comes to precision medicine, discovering a potentially lengthy rule that is specific to a patient could be of interest. In addition, the choice of how many causal rules to discover in the discovery step may depend on the questions that practitioners want to answer. For example, policymakers generally want to discover a short list of risk factors. Some important subgroups defined by risk factors are usually easy to understand and further foster focused discussions about the assessments of potential risks and benefits of policy actions. With limited resources, public health can be effectively promoted when vulnerable subgroups are known and prioritized. In contrast, in precision medicine, a comparatively larger set of decision rules can be chosen. In fact, an important goal is to identify subgroups of patients who respond to treatment at a much higher (or lower) rate than the average \citep{loh2019subgroup}. In addition, identifying a subgroup that must avoid treatment due to its excessive side effects can be valuable information. However, discovering only a few subgroups is likely to miss this extreme subgroup. 

From simulations, we showed that CRE has competitive performance both in the discovery and estimation of the treatment effect. We showed that under the model in equation \eqref{eq:catelindecomposition} and the notable heterogeneity in the causal effects, CRE perfectly retrieves the correct treatment effect decomposition and outperforms the corresponding standalone pseudo-outcome estimators. All simulation studies are repeated with different data-generating processes, leading to coherent results.

The use of CRE allowed for the identification of crucial factors in characteristics that help explain the varying levels of susceptibility to air pollution exposure in the elderly population in the U.S. By employing CRE, a nontrivial and comprehensible characterization of distinct subgroups has been retrieved. This application not only showcased the efficacy of CRE in deciphering complex patterns in data, but also highlighted the importance of understanding the heterogeneous nature of populations in relation to environmental hazards. Such insights should be paired with extensive analyses of vulnerability to air pollution in the younger population. The results of our analyses found indeed that there may be room for possible survival bias when analyzing the vulnerability to air pollution due to structural differences in air pollution exposure across the U.S.

A ubiquitous limitation of observational studies is the possibility of unmeasured confounding bias. Our proposed CRE method can be supplemented with existing approaches for conducting sensitivity analyses to unmeasured confounding bias in causal inference, as we have outlined in Subsection \ref{subsec:confounding}. A similar approach was proposed in our previous work in the context of HTE discovery using a single-tree algorithm and similar data \citep{lee2021discovering}. Despite single-tree algorithms being more prone to discovering spurious HTE, the majority of the discovered subgroups were found to be robust against unmeasured confounding bias. This suggests that our case could be similar. However, we propose extending the sensitivity analyses for confounding bias in the context of our proposed CRE algorithm as a promising direction for future research.

Several other extensions of the CRE method can be possible. CRE deals with the exploration of heterogeneous treatment effects in the case of a binary treatment in a cross-sectional setting. It would be of great interest to extend the CRE setting to continuous treatment effects and time series studies, as these dimensions might be critically important for a number of applications in social and health sciences. In this paper, we have considered a binary threshold for air pollution (i.e., levels of PM$_{2.5}$ being greater than 12 $\mu g \slash m^3$ or not). As we specified, we chose to binarize our exposure because this threshold is highly relevant from a policy perspective, as it was the NAAQS established by the EPA during our study period. However, there may be situations in which the interest lies in looking at HTE along the entire PM$_{2.5}$ exposure response curve. 

Furthermore, starting from CRE to develop interpretable methods for optimal policies or the assignment of targeted treatment is also crucial. The purpose of this paper is to provide a tool for the discovery of HTE. However, a clear next step would be to develop an interpretable tool for optimal policies. Optimal policies involve assigning treatments to individuals in a manner that maximizes the desired outcome while taking into account their unique characteristics. Such an approach could lead to more effective and efficient treatment outcomes, reduce unnecessary treatment, and improve patient outcomes. Additionally, by targeting treatments to the appropriate individuals, optimal policies can help reduce health disparities and ensure that interventions are more equitably distributed. Therefore, developing optimal policies methods is an important future extension of CRE in the direction of moving one more step closer to achieving personalized and effective healthcare.

%% file: Appendix/appendix_a.tex
\section{Models used for Pseudo-Outcomes Estimation}
\label{ssec:ite_overview}

We provide here a brief overview of the six pseudo-outcome estimators, highlighting and comparing their strengths and weaknesses. An empirical comparison among these methods is reported in section \ref{sec:simulations}. It is important to note that the methods presented are not an exhaustive list but rather a selection of approaches employed for this task.

\subsection{T-Learner} 

The T-Learner \citepApp{hansotia2002incremental} is a two-step approach where the conditional mean functions:
\begin{align}
    \mu_0(\bx) := \E\left[Y_i(0)\mid\bX_i=\bx\right] \\
    \mu_1(\bx) := \E\left[Y_i(1)\mid\bX_i=\bx\right]
\end{align}
are estimated separately with any supervised learning algorithm (e.g., Generalized Linear Model, Random Forest, XGBoost, Neural Network, and so on).

In the first step, the conditional mean under control is estimated by all the observations in the control group ($\hat{\mu_0}$),  and the conditional mean under treatment is estimated by all the observations in the treated group ($\hat{\mu_1}$).
Then, exploiting equation \eqref{eq:CATEidentification}, the Treatment Effect is estimated by:
\begin{equation}
    \hat{\tau}(\bx) =  \hat{\mu}_1(\bx) - \hat{\mu}_0(\bx). 
\end{equation}

\subsection{S-Learner} 

The S-learner \citepApp{hill2011bayesian} treats the treatment variable $Z_i$ as if it was just another covariate like those in the vector $\bX_i$. Instead of having two models for the response as a function of the covariates, the S-learner has a single model for the response as a function of the covariates and the treatment:
\begin{equation}
    \mu(\bx, z) := \E\left[Y_i\mid\bX_i=\bx, Z_i = z\right].
\end{equation}

In the first step, all the observations are used to estimate the response function above, $\hat{\mu}$, by any supervised learning algorithm (e.g., Generalized Linear Model, Tree Ensemble, Neural Network). Then, exploiting equation \eqref{eq:CATEidentification}, the Treatment Effect is estimated by:
\begin{equation}
\label{eq:slearner}
    \hat{\tau}(\bx) =  \hat{\mu}(\bx, 1) - \hat{\mu}(\bx, 0).
\end{equation}

Two popular S-Learner variants are the Causal Bayesian Additive Regression Trees (Causal BART) approach \citepApp{hill2011bayesian} and the Bayesian Causal Forest (BCF) approach \citepApp{hahn2020bayesian}, both relying on Bayesian Additive Regression Trees (BART) approach \citepApp{chipman2010bart}, which combines gradient-boosting trees in a Bayesian framing using Markov Chain Monte Carlo (MCMC) sampling for back fitting (using additive and generalized additive models for posterior sampling). 

%\paragraph{Causal Bayesian Additive Regression Trees (Causal BART)}
Causal BART relies on such non-parametric Bayesian models to estimate treatment effects via S-Learner estimator by equation \eqref{eq:slearner}. The method is designed to estimate the treatment effect from observational studies with small effect sizes and heterogeneous effects.

%\paragraph{Bayesian Causal Forest (BCF)} 
BCF combines Bayesian regularization with regression trees to provide a highly flexible response surface that, thanks to regularization from prior distributions, does not overfit the training data. In particular, BCF models the response as a function of the covariates and the treatment, adding the following priors:
\begin{align}
    \mu(\bx, z) :&= \E\left[Y_i\mid\bX_i=\bx, Z_i = z\right] \\
    &=  \mu(\bx, \hat{e}(z)) + \tau(\bx)z.
\end{align}
where $\hat{e}$ is the estimated propensity score and the functions $\mu$ and $\tau$ are independent BART priors. The inclusion of the estimated propensity score can be seen as a covariate-dependent prior to controlling for confounding bias. The treatment effect is then computed as an S-Learner estimator by equation \eqref{eq:slearner}. 

In the rest of the paper, we use S-Learner to refer to the vanilla formulation with a classic supervised learning estimator for $\mu$, e.g., XGboost \citepApp{chen2016xgboost}, and CausalBART and BCF to refer to these two specific variants.

\subsection{X-Learner} 

The X-learner \citepApp{kunzel2019metalearners} is a three steps approach, estimating a treatment effect separately for the control and the treatment group. In the first step the conditional mean functions:
\begin{align}
    \mu_0(\bx) := \E\left[Y_i(0)\mid\bX_i=\bx\right] \\
    \mu_1(\bx) := \E\left[Y_i(1)\mid\bX_i=\bx\right]
\end{align}
are estimated separately by any supervised learning algorithm (e.g., Generalized Linear Model, Random Forest, Neural Network).

The conditional mean under control is estimated by all the observations in the control group ($\hat{\mu_0}$), and the conditional mean under treatment is estimated by all the observations in the treated group ($\hat{\mu_1}$). Secondly, these two estimates are used for predicting the counterfactual outcomes.
\begin{align}
    \hat{\Psi}_1(\bX_i) = Y_i - \hat{\mu_0}(\bX_i) \qquad \text{ for } i: Z_i = 1, \\
\hat{\Psi}_0(\bX_i) = \hat{\mu_1}(\bX_i) - Y_i \qquad \text{ for } i: Z_i = 0.
\end{align}

Finally, these imputed effects are regressed individually on the covariates to obtain $\hat{\tau}_0$ (the CATE for the control group) and $\hat{\tau}_1$ (the CATE for the treatment group), and then combined by a weight function $g \in [0,1]$:
\begin{equation}
\hat{\tau}(\bx) = g(\bx)\hat{\tau}_0(\bx) + [1-g(\bx)] \hat{\tau}_1(\bx).
\end{equation}
A suitable choice for $g$ is an estimate of the propensity score.

\subsection{Augmented Inverse Probability Weighting (AIPW)} 

The Augmented Inverse Probability Weighting estimator \citepApp{robins1994estimation,robins1997toward} extends the observations balancing of Inverse Probability Weighting methods \citepApp{horvitz1952generalization} with conditional response estimation of the S-Learner, inheriting the benefits of both the approaches. 

Firstly, the conditional mean response:
\begin{equation}
    \mu(\bx, z) := \E\left[Y_i\mid\bX_i=\bx, Z_i = z\right]
\end{equation}
is estimated from all the observations by any supervised learning algorithm, $\hat{\mu}$ (first step S-Learner).
Then, the propensity score:
\begin{equation}
    e(\bx) := \E\left[Z_i\mid\bm{X}_i = \bm{x}\right]
\end{equation}
is estimated from all the observations by any supervised learning algorithm, $\hat{e}(\bx)$.

Finally, the treatment effect is computed by:
\begin{equation}
\label{eq:aipw}
\begin{split}
    \hat{\tau}(\bx) = \frac{1}{|\mathcal{I}(\bx)|} \sum_{i \in \mathcal{I}(\bx)} \bigg\{ &\left( \hat{\mu}(\bX_i,1) + \frac{Z_i(Y_i - \hat{\mu}(\bX_i,1))}{\hat{e}(\bX_i)} \right)  \\
    - & \left(\hat{\mu}(\bX_i,0) + \frac{(1 - Z_i)(Y_i - \hat{\mu}(\bX_i, 0))}{1 - \hat{e}(\bX_i)} \right) \bigg\}
\end{split}
\end{equation}
where $\mathcal{I}(\bx) = \{i \in \mathcal{I}: \bX_i = \bx\}$. By construction, it is only required that one estimator among $\hat{e}$ and $\hat{\mu}$ is unbiased in order to get an unbiased estimate of the treatment effect.

% \subsection{Causal Forest (CF)} 

% The Causal Forest is a method from Generalized Random Forests \citepApp{athey2019generalized}. Similarly to Random Forest \citepApp{breiman2001random}, Causal Forest attempts to find neighborhoods in the covariate space (recursive partitioning). While a Random Forest is built from Decision Trees, a Causal Forest is built from Causal Trees \citepApp{athey2016}, which splitting criterion optimizes for finding splits associated with treatment effect heterogeneity. The goal is to find leaves where the treatment effect is constant but is different from other leaves. 

% A Causal Forest is simply the average of a large number of Causal Trees, where the trees differ due to subsampling. To create a Causal Forest from Causal Trees, it is necessary to estimate a weighting function and use the resulting weights to solve a local generalized method of moments (GMM) model to estimate the CATE. To deal with overfitting and biased estimations, Causal Forests, like Causal Rule Ensemble itself, rely on the honesty condition, whereby each training sample $i$ is only used to decide where to place the split (discovery) or to estimate the within-leaf treatment effect (estimation), but not both. Honesty condition also leads to asymptotic normality.
% The prediction of treatment effects is the difference in the average outcomes between the treated and the control observations of the estimating subsample in terminal leaves.

%% file: Appendix/appendix_a1.tex
\section{Filtering}

In the discovery step, once a set of candidate decision rules ($\hat{\mathcal{R}}''$) has been proposed, we discard all the extreme or redundant decision rules based on the following two criteria:

\begin{enumerate}[i.]
    \item \textbf{Extreme}: A (candidate) decision rule $r_m \in \hat{\mathcal{R}}''$ is said extreme if either too rare:
    \begin{equation}
    \sum_{i \in \mathcal{I}^{dis}} r_m(\bX_i) \leq t_{ext}N^{dis},
    \end{equation}
    or too common:
    \begin{equation}
    \sum_{i \in \mathcal{I}^{dis}} r_m(\bX_i) \geq (1-t_{ext}) N^{dis},
    \end{equation}
    where $t_{ext}$ is the threshold parameter defining the limit ratio, and $N^{dis} = |\mathcal{I}^{dis}|$. 
    \item \textbf{Redundant}: A (candidate) decision rule $r_a \in \hat{\mathcal{R}}''$ is said redundant if exists at least another (candidate) decision rule $r_b \in \hat{\mathcal{R}}''$ such that their correlation is greater than a fixed threshold ($t_{corr}$).
\end{enumerate}

These filtering criteria help in practice to preliminary filter irrelevant decision rules and speed up the rules selection step.

%% file: Appendix/appendix_b.tex
\section{Mathematical Proofs}
\label{app:extrathoery}

We report here the proofs of the propositions presented in section \ref{sec:method}.

\subsection{Proposition 1. (Consistency of the AATE estimator)}

\paragraph{Proposition 1.}
\textit{Let $\hat{\bm{\tau}}$ a consistent estimator for $\bm{\tau}$ (i.e., AIPW). Let $\E(\bm{R}^T \bm{R}) = Q$ a positive definite matrix, then the CRE estimator defined in equation \eqref{eq:OLSestimator} is a consistent estimator for $\bm{\alpha}$ (AATE).}

\begin{proof}
Multiplying equation \eqref{eq:mlindecomposition} on the both sides by $(\bm{R}^T \bm{R})^{-1} \bm{R}^T$, we get:
\begin{equation}
    (\bm{R}^T \bm{R})^{-1} \bm{R}^T(\hat{\bm{\tau}} - \hat{\bar{\tau}}) =  (\bm{R}^T \bm{R})^{-1} \bm{R}^T\bm{R} \bm{\alpha} + (\bm{R}^T \bm{R})^{-1} \bm{R}^T\bm{\nu}.
\end{equation}
Using equation \eqref{eq:OLSestimator}, and simplifying the right member:
\begin{equation}
\label{eq:preconsistencyAATE}
    \rightarrow \quad \hat{\bm{\alpha}} =  \bm{\alpha} + (\bm{R}^T \bm{R})^{-1} \bm{R}^T\bm{\nu}.
\end{equation}
By the Law of large numbers:
\begin{equation}
    \frac{\bm{R}^T \bm{R}}{N} = \frac{1}{N}\sum_{i=1}^N \bm{\bm{R}}_i^T \bm{\bm{R}}_i \overset{d}{\to} \bm{Q} \succ \bm{0},
\end{equation}
and:
\begin{equation}
    \frac{\bm{R}^T \bm{\nu}}{N} = \frac{1}{N}\sum_{i=1}^N \bm{R}_i^T \cdot \nu_i \overset{d}{\to} \bm{0},
\end{equation}
where $\bm{R}_i$ represents the $i$-th row of the rules matrix.
Combining these results in equation \eqref{eq:preconsistencyAATE} (simplifying $N$), by Slutsky's theorem:
\begin{equation}
    \quad \hat{\bm{\alpha}} \overset{d}{\to}  \bm{\alpha}
\end{equation}
\end{proof}

\subsubsection{Proposition 2. (Asymptotic Normality of the AATE estimator)}

\paragraph{Proposition 2.}
%\label{prop:alpha_asnormality}
\textit{Under Proposition \eqref{prop:consistency} assumptions, and let $\E(\nu_i^4) < \infty$ and $\E(\nu_i^2 \bm{R}_i^T \bm{R}_i) = \Omega \succ 0$ (positive definite) for all $i \in \mathcal{I}^{inf}$ (additional regularity assumptions),
then:
\begin{equation}
    \sqrt{N} (\hat{\bm{\alpha}} - \bm{\alpha})  \overset{d}{\to}  \mathcal{N}(\bm{0}, \bm{V}) \:\:\:  \text{as} \:\:\: N \to \infty
\end{equation}
where $\bm{R}_i$ represents the $i$-th row of the rules matrix $\bm{R}$ and $\bm{V} = \bm{Q}^{-1} \bm{\Omega} \bm{Q}^{-1}$.}

\begin{proof}
Similarly to the proof of Proposition \eqref{prop:consistency}, multiplying equation \eqref{eq:mlindecomposition} on the both sides by $(\bm{R}^T \bm{R})^{-1} \bm{R}^T$, and inserting equation \eqref{eq:OLSestimator}, we get:
\begin{equation}
\label{eq:preconsistencyAATE2}
 \hat{\bm{\alpha}} =  \bm{\alpha} + (\bm{R}^T \bm{R})^{-1} \bm{R}^T\bm{\nu}.
\end{equation}
Multiplying both sides by $\sqrt{N}$ and rearranging we get:
\begin{equation}
\label{eq:prenormalityAATE}
\begin{split}
    \sqrt{N} (\hat{\bm{\alpha}}-\bm{\alpha}) &=  \left(\frac{\bm{R}^T \bm{R}}{N}\right)^{-1} \frac{\bm{R}^T\bm{\nu}}{\sqrt{N}} \\
     &= \left(\frac{\sum_{i=1}^N \bm{R}_i^T \bm{R}_i}{N}\right)^{-1} \frac{\sum_{i=1}^N \bm{R}_i^T\nu_i}{\sqrt{N}} \\
\end{split}
\end{equation}
By hypothesis:
\begin{equation}
    \E[\bm{R}_i\nu_i] = \bm{0} \qquad \forall i \in \mathcal{I}^e
\end{equation}
and:
\begin{equation}
\begin{split}
    \text{Var}(\bm{R}_i\nu_i) 
    &= \E[\nu_i^2\bm{R}_i^T\bm{R}_i] - \E[\bm{R}_i\nu_i]^2 \\
    &= \E[\nu_i^2\bm{R}_i^T\bm{R}_i] = \bm{\Omega} \succ \bm{0} \qquad \forall i \in \mathcal{I}^e
\end{split}
\end{equation}
Then, by the Central Limit Theorem:
\begin{equation}
    \frac{\sum_{i=1}^N \bm{R}_i^T \nu_i}{N} \overset{d}{\to}  \mathcal{N}(\bm{0}, \bm{\Omega}).
\end{equation}
We have already discussed in the proof of Proposition \eqref{prop:consistency} that:
\begin{equation}
    \frac{\bm{R}^T \bm{R}}{N} = \frac{1}{N}\sum_{i=1}^N \bm{R}_i^T \cdot \bm{R}_i \overset{d}{\to} \bm{Q} \succ \bm{0}.
\end{equation}
Then, combining these results in equation \eqref{eq:prenormalityAATE}, by Slutsky' theorem and Cramer-Wold theorem: 
\begin{equation}
    \sqrt{N} (\hat{\bm{\alpha}} - \bm{\alpha})  \overset{d}{\to}  \mathcal{N}(\bm{0}, \bm{Q}^{-1} \bm{\Omega} \bm{Q}^{-1}) \:\:\:  \text{as} \:\:\: N \to \infty
\end{equation}
\end{proof}

%% file: Appendix/appendix_c.tex
\section{Additional Simulations}
\label{app:extrasimulations}

We present here a more extensive analysis of the heterogeneity discovery and treatment effect estimation simulation studies under different variants of the data-generating process, varying sample size, the number of decision rules, the complexity of the decision rules, and the type of confounding. In particular, for both the simulation studies, we consider the following five variants to the data generating process described in section \ref{sec:simulations} (where all the definitions are kept equal if not otherwise specified):
\begin{enumerate}[i.]
    \item \textbf{Large Sample}: $N=5,000$ individuals, $M=2$ rules ($r_1$, $r_2$), linear confounding;
    \item \textbf{Small Sample}: $N=1,000$ individuals, $M=2$ rules ($r_1$, $r_2$), linear confounding;
    \item \textbf{More Rules}: $N=2,000$ individuals, $M=4$ rules ($r_1$, $r_2$, $r_3$, $r_4$), linear confounding, where:
    \begin{eqnarray*}
        \mu_i^0 &=& f(\bm{X}_i)+k \cdot \mathds{1}_{\{x_1=1;x_2=0\}}(\bm{X}_i) +\frac{k}{2} \cdot \mathds{1}_{\{x_4=0\}}(\bm{X}_i) \\
         &=& f(\bm{X}_i)+k \cdot r_1(\bm{X}_i) +\frac{k}{2} \cdot r_3(\bm{X}_i), \nonumber
    \end{eqnarray*}
    \begin{eqnarray*}
      \mu_i^1 &=& f(\bm{X}_i)+k \cdot \mathds{1}_{\{x_5=1;x_6=0\}}(\bm{X}_i)+2k \cdot \mathds{1}_{\{x_5=0;x_7=1;x_8=0\}}(\bm{X}_i) \\
        &=& f(\bm{X}_i)+k \cdot r_2(\bm{X}_i)+2k \cdot r_4(\bm{X}_i),  \nonumber
    \end{eqnarray*}
    and then:
    \begin{equation*}
        \tau(\bm{x}) = - k \cdot r_1(\bm{x}) + k \cdot r_2(\bm{x}) - \frac{k}{2} \cdot r_3(\bm{x}) + 2k \cdot r_4(\bm{x}) ; 
    \end{equation*}
    \item (Pseudo) \textbf{Randomized Controlled Trial}: $N=2,000$ individuals, $M=2$ rules ($r_1$, $r_2$), no confounding, only effect modification by the decision rules, i.e.:
    \begin{eqnarray*}
        \mu_i^0 &= k \cdot \mathds{1}_{\{x_1=1;x_2=0\}}(\bm{X}_i) = k \cdot r_1(\bm{X}_i), \\
        \mu_i^1 &= k \cdot \mathds{1}_{\{x_5=1;x_6=0\}}(\bm{X}_i) = k \cdot r_2(\bm{X}_i);
    \end{eqnarray*}
    and then (as the original data-generating process):
    \begin{equation*}
        \tau(\bm{x}) = - k \cdot r_1(\bm{x}) + k \cdot r_2(\bm{x}); 
    \end{equation*}
    \item \textbf{Non-Linear Confounding}: $N=2,000$ individuals, $M=2$ rules ($r_1$, $r_2$), non-linear confounding, i.e.:
    \begin{align*}
        \mu_i^0 &= g(\bm{X}_i) + k \cdot \mathds{1}_{\{x_1=1;x_2=0\}}(\bm{X}_i) = g(\bm{X}_i) +  k \cdot r_1(\bm{X}_i), \\
        \mu_i^1 &= g(\bm{X}_i) + k \cdot \mathds{1}_{\{x_5=1;x_6=0\}}(\bm{X}_i) = g(\bm{X}_i) + k \cdot r_2(\bm{X}_i),
    \end{align*}
    where:
    \begin{equation*}
        g(\bm{X}_i) = X_i^1 + \cos(X_i^3 \cdot X_i^4);
    \end{equation*}
    and then (as the original data-generating process):
    \begin{equation*}
        \tau(\bm{x})= - k \cdot r_1(\bm{x}) + k \cdot r_2(\bm{x}).
    \end{equation*}
\end{enumerate}

Each of the described data-generating processes vary the original design for a specific characteristic, which we desire to test our methodology on. In subsection \ref{sec:app_exp_discovery} we report and discuss the results of the heterogeneity discovery simulation study over these different data-generating processes, and in subsection \ref{sec:app_exp_estimation}, we report and discuss the results of the heterogeneous treatment effect estimation simulation study over the same instances, also including empirical verification of the consistency of the AATE estimation.

\subsection{Discovery}
\label{sec:app_exp_discovery}

In this section, we discuss, one by one, the results of the simulations study on heterogeneity discovery presented in section \ref{sec:sym_discovery} on the five variant data-generating processes described above. 

\subsubsection{Large Sample}
\begin{figure}[h]
	\centering
	\includegraphics[width=\textwidth]{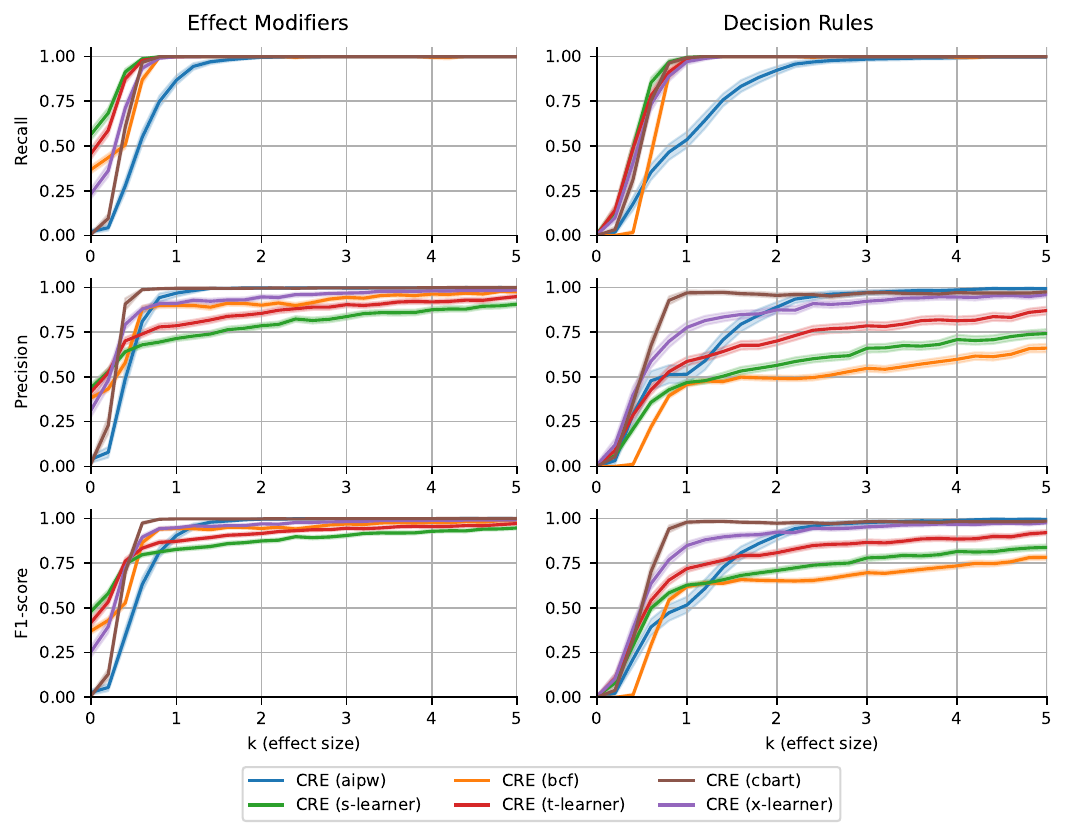} 
	\caption{Simulation study for heterogeneity discovery results with 2 rules, linear confounders and 5,000 observations. Mean $Precision$, $Recall$ and $F1-score$ (lines) with the corresponding 95\% confidence intervals (bands) over 250 Monte Carlo experiments are reported for each method and causal effect size $k$. %For each CRE variant, the heterogeneity characterization discovery converges (with respect to effect size) to the true heterogeneity characterization.
 }
	\label{fig:discovery_big_sample}
\end{figure}

In Figure \ref{fig:discovery_big_sample} we report the results for heterogeneity discovery increasing the sample size to $N=5,000$ individuals. 
As expected, all the methods follow the same trends described for the original data-generating process, but significantly increase the convergence rate, in particular for the $Recall$ in both Estimation and Decision Rules retrieval. %CRE (AIPW) is the unique method not speeding up the convergence rate to perfect recovery, probably due to its instability issued already discussed in section \ref{sec:sym_estimation}.

\subsubsection{Small Sample}

\begin{figure}[h]
	\centering
	\includegraphics[width=\textwidth]{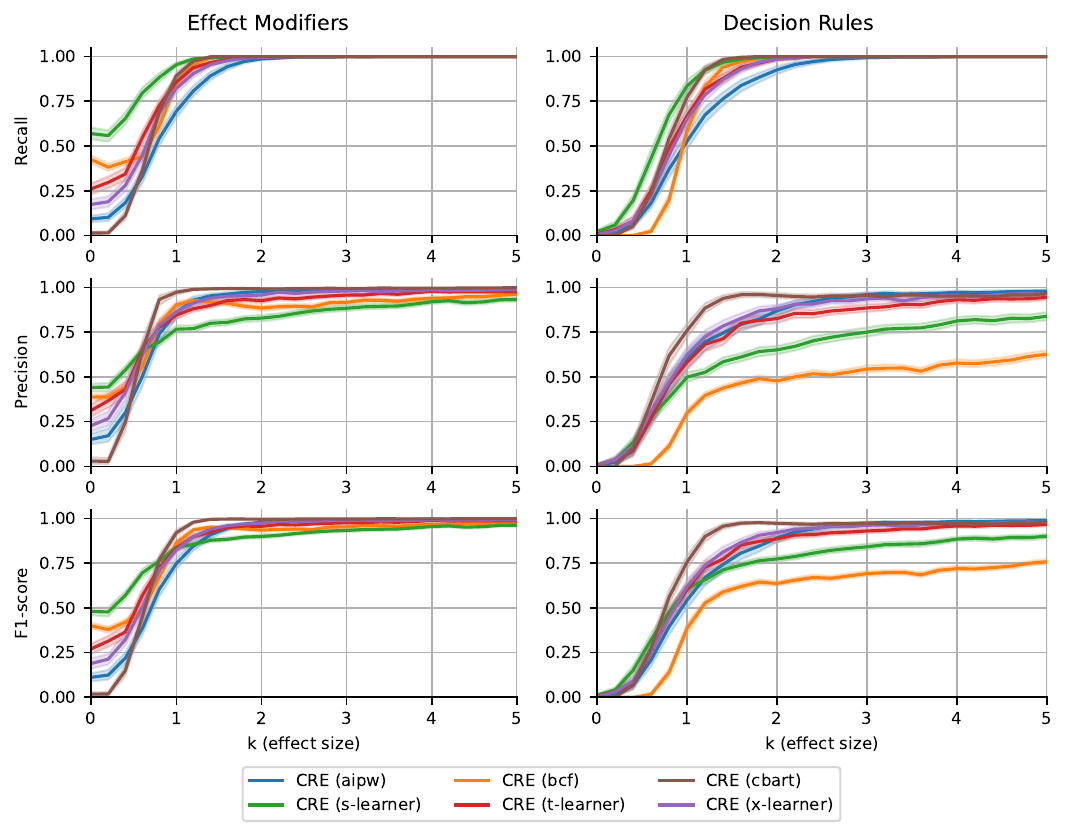} 
	\caption{Simulation study for heterogeneity discovery results with 2 rules, linear confounders, and 1,000 observations. Mean $Precision$, $Recall$, and $F1-score$ (lines) with the corresponding 95\% confidence intervals (bands) over 250 Monte Carlo experiments are reported for each method and causal effect size $k$. %For each CRE variant, the heterogeneity characterization discovery converges (with respect to effect size) to the true heterogeneity characterization.
 }
	\label{fig:discovery_small_sample}
\end{figure}

In Figure \ref{fig:discovery_small_sample} we report the results for heterogeneity discovery decreasing the sample size to $N=1,000$ individuals. 
As expected, all the methods follow the same trends described for the original data-generating process, without significantly decreasing the convergence rate toward perfect discovery. These results are indicative of a good performance of CRE even in smaller sample regimes.

\subsubsection{More Rules}

\begin{figure}[h]
	\centering
	\includegraphics[width=\textwidth]{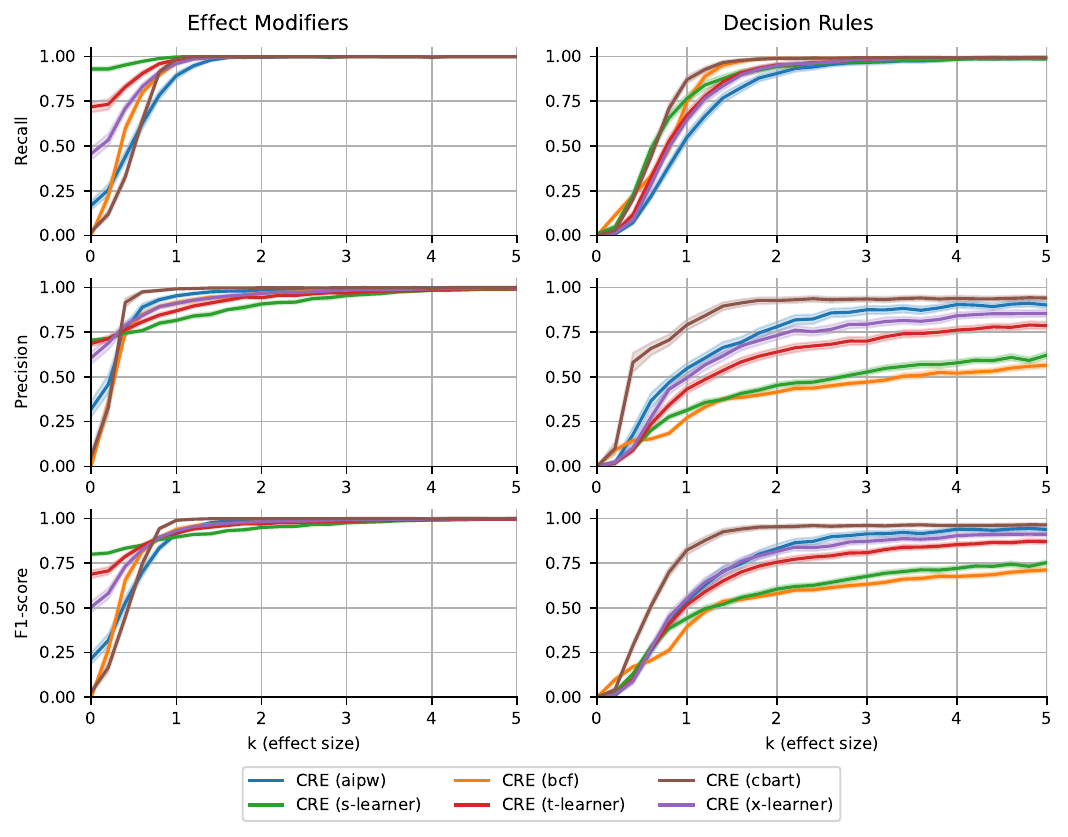} 
	\caption{Simulation study for heterogeneity discovery results with 4 rules, linear confounders, and 2,000 observations. Mean $Precision$, $Recall$ and $F1-score$ (lines) with the corresponding 95\% confidence intervals (bands) over 250 Monte Carlo experiments are reported for each method and causal effect size $k$. %For each CRE variant, the heterogeneity characterization discovery converges (with respect to effect size) to the true heterogeneity characterization.
 }
	\label{fig:discovery_more_rules}
\end{figure}

In Figure \ref{fig:discovery_more_rules}, we report the results for heterogeneity discovery, increasing the number of decision rules to $M=4$.
As expected, (almost) all the methods increase their discovery performances, increasing the causal effect ($k$). There are now seven effect modifiers out of $p=10$ covariates, which leads to easier effect modifiers retrieval. The decision rules discovery is instead more challenging due to the higher and heterogeneous/more complex number of rules to retrieve. All the methods still perfectly retrieve all the true decision rules with $k>3$ ($Recall=1$), but they more often retrieve also wrong or redundant rules ($Precision<1$).
%CRE (CF) is the unique method that does not show a significant dependence on the causal effect for $k>1$. Our hypothesis is that during the ITE estimation, it struggles more than the other methods in trying to express the heterogeneity in the longest rules (i.e., $r_3$) through its causal trees.

\subsubsection{Randomized Controlled Trial}

\begin{figure}[h]
	\centering
	\includegraphics[width=\textwidth]{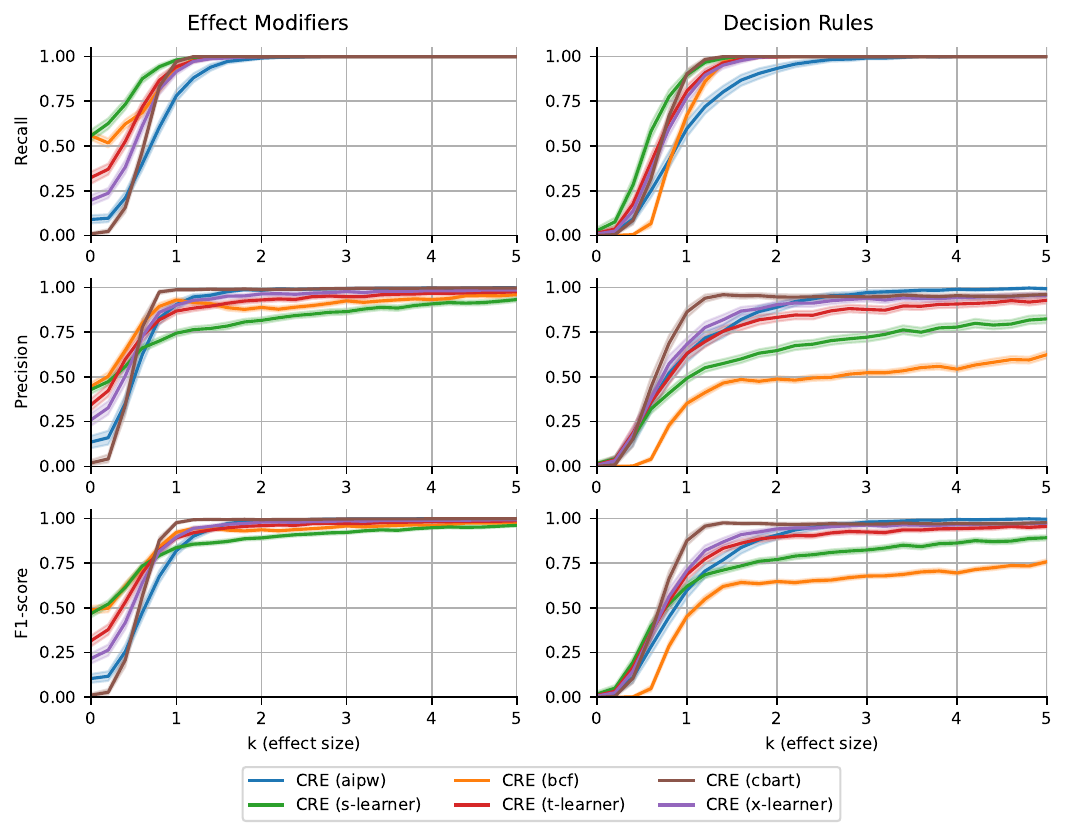} 
	\caption{Simulation study for heterogeneity discovery results with 2 rules, no confounders (randomized controlled trial), and 2,000 observations. Mean $Precision$, $Recall$, and $F1-score$ (lines) with the corresponding 95\% confidence intervals (bands) over 250 Monte Carlo experiments are reported for each method and causal effect size $k$. %For each CRE variant, the heterogeneity characterization discovery converges (with respect to effect size) to the true heterogeneity characterization.
 }
	\label{fig:discovery_rct}
\end{figure}

In Figure \ref{fig:discovery_rct}, we report the results for heterogeneity discovery with no confounding.
As expected, all the methods follow the same trends described for the original data-generating process, but significantly and equally increase the convergence rate, in particular for the $Recall$ in both estimation and decision rules retrieval. Indeed, removing additional confounding mechanism facilitate all the estimation steps.

\subsubsection{Non-Linear Confounding}
\begin{figure}[h]
	\centering
	\includegraphics[width=\textwidth]{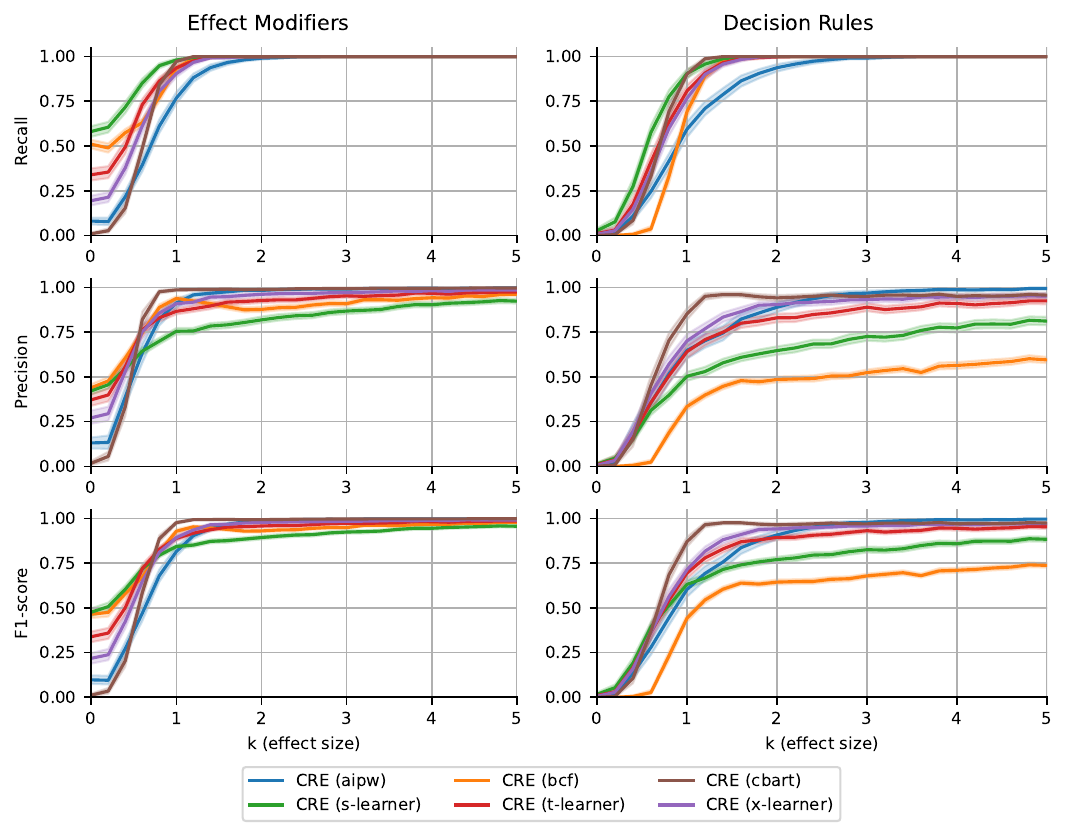} 
	\caption{Simulation study for heterogeneity discovery results with 2 rules, non-linear confounders, and 2,000 observations. Mean $Precision$, $Recall$, and $F1-score$ (lines) with the corresponding 95\% confidence intervals (bands) over 250 Monte Carlo experiments are reported for each method and causal effect size $k$. %For each CRE variant, the heterogeneity characterization discovery converges (with respect to effect size) to the true heterogeneity characterization.
 }
	\label{fig:discovery_nonlin}
\end{figure}

In Figure \ref{fig:discovery_nonlin}, we report the results for heterogeneity discovery with non-linear confounding.
As expected, all the methods follow the same trends described for the original data-generating process, without significantly decreasing the convergence rate towards perfect discovery, although the more complex confounding mechanism.

\subsection{Estimation}
\label{sec:app_exp_estimation}
In this section, we discuss, one by one, the results of the simulations study on heterogeneous treatment effect estimation presented in section \ref{sec:sym_estimation} on the five variant data generating processes described above. In addition, we show the empirical consistency of the AATE estimation by reporting a boxplot of the AATEs estimation bias $$\text{Bias}(r_m) = \alpha_m - \hat{\alpha}_m \qquad r_m \in \mathcal{R}$$ for all the CRE variants, for all the data generating process variants including the main setup described in section \ref{sec:sym_estimation}. 

\subsubsection{Main}
\begin{figure}[h!]
	\centering
	\includegraphics[width=\textwidth]{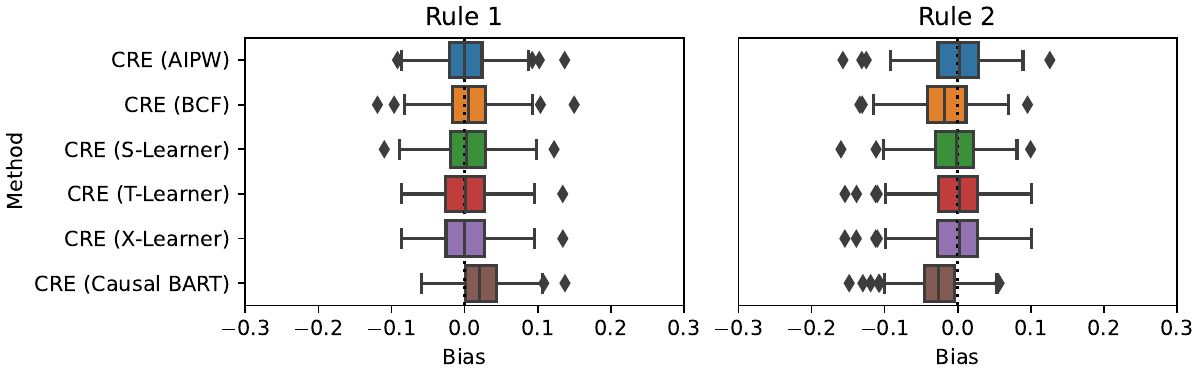} 
	\caption{Simulation study for HTE estimation, with $M=2$ rules, linear confounding, and 2,000 individuals. For all the CRE variants, for each rule, the AATE's bias over 250 Monte Carlo experiments is reported in a boxplot.}
	\label{fig:alpha_main}
\end{figure}

In Figure \ref{fig:alpha_main}, we report a boxplot of the AATEs estimation bias  comparing the different CRE variants over the same 250 Monte Carlo experiments summarized in Table \ref{tab:estimation_main}.
As expected from Proposition \ref{prop:consistency}, all the CRE variants lead to consistent AATEs estimation (median centered in 0), even without assuming perfect rules discovery.

\subsubsection{Large Sample}
In Table \ref{tab:estimation_big_sample}, we report the results for heterogeneous treatment effect estimation, increasing the sample size to $N=5,000$ individuals.
\begin{table}[h!]
\centering
\begin{tabular}{crrrr}
\hline
 & \multicolumn{2}{c}{RMSE} & \multicolumn{2}{c}{Bias} \\
Method & \multicolumn{1}{c}{$\mu$} & \multicolumn{1}{c}{$\sigma$} & \multicolumn{1}{c}{$\mu$} & \multicolumn{1}{c}{$\sigma$} \\ \midrule \midrule
CRE (AIPW) & 0.0804 & 0.0357 & 0.0011 & 0.0527 \\
\rowcolor{gray!10} CRE (BCF) & 0.0891 & 0.0324 & 0.0022 & 0.0488 \\
CRE (S-Learner) & 0.0918 & 0.0332 & 0.0010 & 0.0497 \\
\rowcolor{gray!10} CRE (T-Learner) & 0.0905 & 0.0343 & 0.0034 & 0.0522 \\
CRE (X-Learner) & 0.0850 & 0.0337 & 0.0034 & 0.0522 \\
\rowcolor{gray!10} CRE (Causal BART) & \textbf{0.0781} & 0.0306 & \textbf{0.0002} & 0.0490 \\
\hline
AIPW & 2.2526 & 0.0910 & 0.0016 & 0.0342 \\
\rowcolor{gray!10}BCF & 0.0814 & 0.0234 & 0.0020 & 0.0319 \\
S-Learner & 0.3110 & 0.0218 & 0.0012 & 0.0333 \\
\rowcolor{gray!10}T-Learner & 0.5090 & 0.0214 & 0.0024 & 0.0333 \\
X-Learner & 1.0756 & 0.0140 & 0.0024 & 0.0333 \\
\rowcolor{gray!10}Causal BART & 0.9977 & 0.0099 & 0.0003 & 0.0315 \\ \hline
\end{tabular}
\caption{Simulation study for HTE estimation, with $M=2$ rules, linear confounder, 5,000 individuals. For all the methods, the mean ($\mu$) and standard deviation ($\sigma$) treatment effect root mean squared error (RMSE) and bias (Bias) over 250 Monte Carlo experiments are reported.}
\label{tab:estimation_big_sample}
\end{table}
As discussed in section \ref{sec:sym_estimation}, all the CRE variants significantly outperform the corresponding `standalone' pseudo-outcome estimators in both pseudo-outcome and ATE estimation. Bayesian Causal Forest is the unique pseudo-outcome estimator in getting similar performances to the corresponding CRE variant. %AIPW estimator still suffers from not stabilized pseudo-outcome prediction. %Causal Forest, and similarly CRE (CF), significantly improve their estimation performances with respect to the original data generating process, leading to unbiased estimation, enforcing our hypothesis of empirically slower Causal Forest consistency convergence rate.

\begin{figure}[h!]
	\centering
	\includegraphics[width=\textwidth]{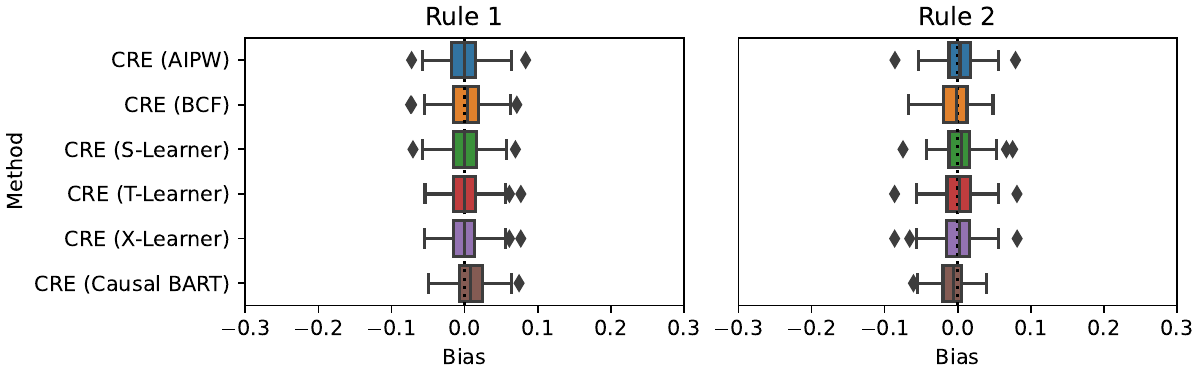}
	\caption{Simulation study for HTE estimation, with $M=2$ rules, linear confounding and 5,000 individuals. For all the CRE variants, for each rule, the AATE's bias over 250 Monte Carlo experiments is reported in a boxplot.}
	\label{fig:alpha_big_sample}
\end{figure}

In Figure \ref{fig:alpha_big_sample}, we report the results for AATEs estimation, increasing the sample size to $N=5,000$ individuals.
As expected from Proposition \ref{prop:alpha_asnormality}, all the methods lead to consistent AATEs estimation, with a confidence interval even smaller than the original data-generating process.

\subsubsection{Small Sample}
In Table \ref{tab:estimation_small_sample}, we report the results for heterogeneous treatment effect estimation, decreasing the sample size to $N=1,000$ individuals.
\begin{table}[h!]
\centering
\begin{tabular}{crrrr}
\hline
 & \multicolumn{2}{c}{RMSE} & \multicolumn{2}{c}{Bias} \\
Method & \multicolumn{1}{c}{$\mu$} & \multicolumn{1}{c}{$\sigma$} & \multicolumn{1}{c}{$\mu$} & \multicolumn{1}{c}{$\sigma$} \\ \midrule \midrule
CRE (AIPW) & 0.2195 & 0.0882 & -0.0136 & 0.1343 \\
\rowcolor{gray!10} CRE (BCF) & 0.2215 & 0.0758 & 0.0055 & 0.1182 \\
CRE (S-Learner) & 0.2312 & 0.0880 & -0.0132 & 0.1321 \\
\rowcolor{gray!10} CRE (T-Learner) & 0.1976 & 0.0859 & -0.0265 & 0.1194 \\
CRE (X-Learner) & \textbf{0.1961} & 0.0854 & -0.0265 & 0.1194 \\
\rowcolor{gray!10} CRE (Causal BART) & 0.2225 & 0.0821 & \textbf{-0.0003} & 0.1187 \\
\hline
AIPW & 1.7045 & 0.1320 & -0.0037 & 0.0843 \\
\rowcolor{gray!10}BCF & 0.2056 & 0.0570 & 0.0022 & 0.0847 \\
S-Learner & 0.6844 & 0.0590 & -0.0030 & 0.0808 \\
\rowcolor{gray!10}T-Learner & 1.1078 & 0.0590 & -0.0064 & 0.0858 \\
X-Learner & 1.3403 & 0.0527 & -0.0064 & 0.0858 \\
\rowcolor{gray!10}Causal BART & 0.9862 & 0.0248 & -0.0007 & 0.0850 \\ \hline
\end{tabular}
\caption{Simulation study for HTE estimation, with $M=2$ rules, linear confounder, 1,000 individuals. For all the methods, the mean ($\mu$) and standard deviation ($\sigma$) treatment effect root mean squared error (RMSE) and bias (Bias) over 250 Monte Carlo experiments are reported.}
\label{tab:estimation_small_sample}
\end{table}
As discussed in section \ref{sec:sym_estimation}, (almost) all the CRE variants significantly outperform the corresponding `standalone' pseudo-outcome estimators in both pseudo-outcome and ATE estimation without significantly worsening the performances from the original data-generating process (with larger sample size), with exceptions of BCF. %Indeed, Causal Forest in a small sample regime leads to even more systematic errors in estimation, which drastically propagate in the corresponding CRE variant.

\begin{figure}[h!]
	\centering
	\includegraphics[width=\textwidth]{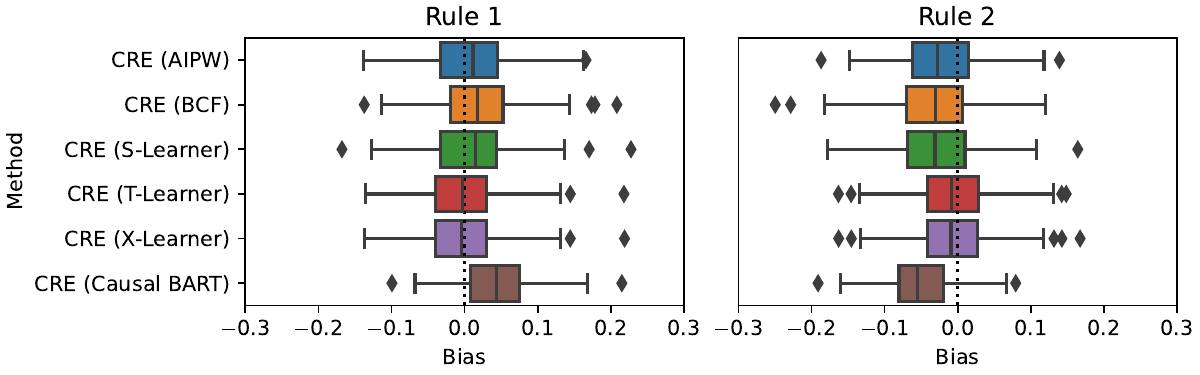} 
	\caption{Simulation study for HTE estimation, with $M=2$ rules, linear confounding and 1,000 individuals. For all the CRE variants, for each rule, the AATE's bias over 250 Monte Carlo experiments is reported in a boxplot.}
	\label{fig:alpha_small_sample}
\end{figure}

In Figure \ref{fig:alpha_small_sample}, we report the results for AATEs estimation, increasing the sample size to $N=1,000$ individuals.
As expected from Proposition \ref{prop:alpha_asnormality}, all the methods lead to consistent AATEs estimation, with no notably larger confidence intervals with respect to the original data generating process, although the sample size.

\subsubsection{More Rules}
In Table \ref{tab:estimation_more_rules}, we report the results for heterogeneous treatment effect estimation, increasing the number of decision rules to $M=4$.
\begin{table}[h!]
\centering
\begin{tabular}{crrrr}
\hline
 & \multicolumn{2}{c}{RMSE} & \multicolumn{2}{c}{Bias} \\
Method & \multicolumn{1}{c}{$\mu$} & \multicolumn{1}{c}{$\sigma$} & \multicolumn{1}{c}{$\mu$} & \multicolumn{1}{c}{$\sigma$} \\ \midrule \midrule
CRE (AIPW) & 0.2505 & 0.1897 & -0.0029 & 0.0936 \\
\rowcolor{gray!10}CRE (BCF) & 0.1901 & 0.0884 & 0.0061 & 0.0804 \\
CRE (S-Learner) & 0.2967 & 0.1394 & -0.0044 & 0.0900 \\
\rowcolor{gray!10}CRE (T-Learner) & 0.2349 & 0.0943 & \textbf{0.0003} & 0.0944 \\
CRE (X-Learner) & 0.2356 & 0.1416 & \textbf{0.0003} & 0.0948 \\
\rowcolor{gray!10}CRE (Causal BART) & 0.1757 & 0.0729 & -0.0017 & 0.0810 \\
\hline
AIPW & 2.1139 & 0.2077 & 0.0007 & 0.0561 \\
\rowcolor{gray!10}BCF & \textbf{0.1698} & 0.0353 & 0.0045 & 0.0519 \\
S-Learner & 0.5410 & 0.0394 & -0.0006 & 0.0532 \\
\rowcolor{gray!10}T-Learner & 0.8075 & 0.0371 & 0.0026 & 0.0567 \\
X-Learner & 1.1883 & 0.0293 & 0.0026 & 0.0567 \\
\rowcolor{gray!10}Causal BART & 0.9994 & 0.0161 & 0.0012 & 0.0517 \\ \hline
\end{tabular}
\caption{Simulation study for HTE estimation, with $M=4$ rules, linear confounder, 2,000 individuals. For all the methods, the mean ($\mu$) and standard deviation ($\sigma$) treatment effect root mean squared error (RMSE) and bias (Bias) over 250 Monte Carlo experiments are reported.}
\label{tab:estimation_more_rules}
\end{table}
As discussed in section \ref{sec:sym_estimation}, (almost) all the CRE variants significantly outperform the corresponding `standalone' pseudo-outcome estimators in both pseudo-outcome and ATE estimation without significantly worsening the performances from the original data-generating process (with simpler CATE decomposition), with exceptions of BCF. %Indeed, Causal Forest in a more complex CATE characterization regime leads to even more systematic errors in estimation, which drastically propagate in the corresponding CRE variant.

\begin{figure}[h!]
	\centering
	\includegraphics[width=0.83\textwidth]{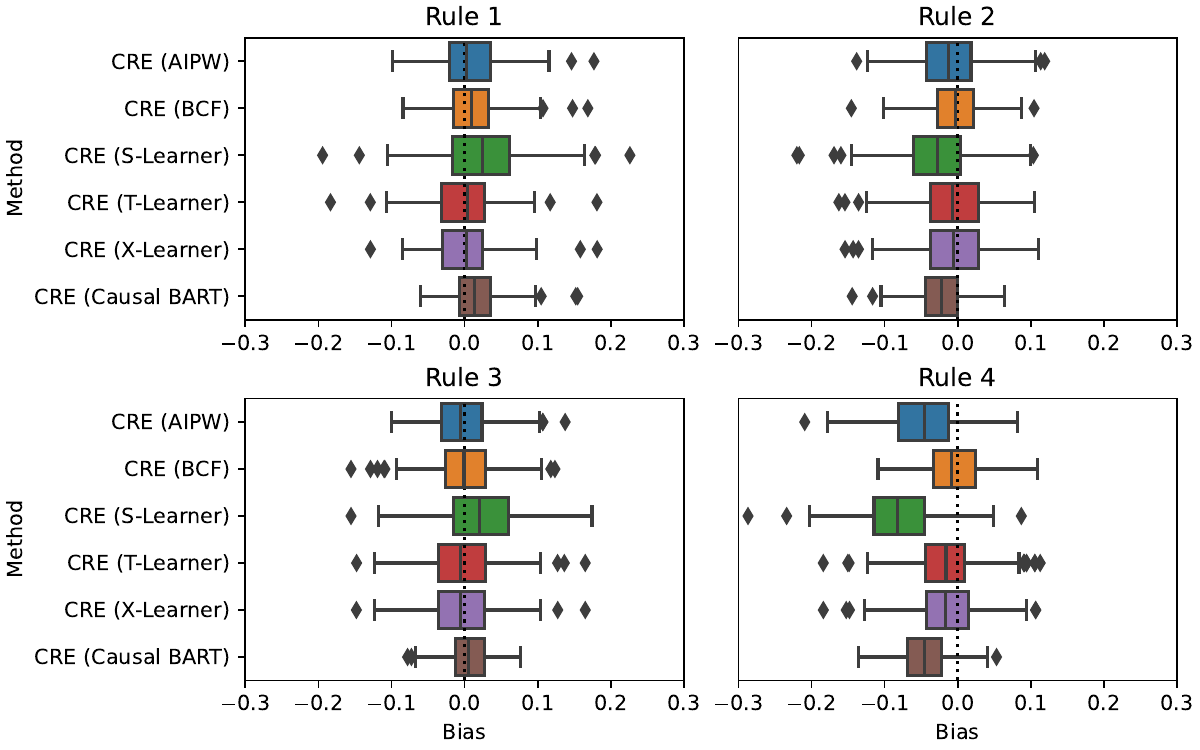}
	\caption{Simulation study for HTE estimation, with $M=4$ rules, linear confounding and 2,000 individuals. For all the CRE variants, for each rule, the AATE's bias over 250 Monte Carlo experiments is reported in a boxplot.}
	\label{fig:alpha_more_rules}
\end{figure}

In Figure \ref{fig:alpha_more_rules}, we report the results for AATEs estimation, increasing the number of decision rules to $M=4$.
As expected from Proposition \ref{prop:alpha_asnormality}, all the methods lead to consistent AATE estimation for (almost) all the rules. Only the fourth and longest rule is  slightly underestimated by almost all the methods, probably due to the redundant recovery of other similar decision rules ($Precision<1$).

\subsubsection{Randomized Controlled Trial}
In Table \ref{tab:estimation_rct}, we report the results for heterogeneous treatment effect estimation, with no confounding (if not in terms of decision rules).
\begin{table}[h!]
\centering
\begin{tabular}{crrrr}
\hline
 & \multicolumn{2}{c}{RMSE} & \multicolumn{2}{c}{Bias} \\
Method & \multicolumn{1}{c}{$\mu$} & \multicolumn{1}{c}{$\sigma$} & \multicolumn{1}{c}{$\mu$} & \multicolumn{1}{c}{$\sigma$} \\ \midrule \midrule
CRE (AIPW) & \textbf{0.1342} & 0.0602 & 0.0031 & 0.0884 \\
\rowcolor{gray!10}CRE (BCF) & 0.1492 & 0.0552 & 0.0047 & 0.0796 \\
CRE (S-Learner) & 0.1465 & 0.0576 & 0.0025 & 0.0848 \\
\rowcolor{gray!10} CRE (T-Learner) & 0.1489 & 0.0659 & 0.0062 & 0.0931 \\
CRE (X-Learner) & 0.1443 & 0.0655 & 0.0062 & 0.0931 \\
\rowcolor{gray!10} CRE (Causal BART) & 0.1457 & 0.0621 & \textbf{0.0010} & 0.0810 \\
\hline
AIPW & 2.0757 & 0.1897 & 0.0039 & 0.0560 \\
\rowcolor{gray!10}BCF & 0.1351 & 0.0371 & 0.0045 & 0.0517 \\
S-Learner & 0.4690 & 0.0342 & 0.0031 & 0.0527 \\
\rowcolor{gray!10}T-Learner & 0.8042 & 0.0367 & 0.0042 & 0.0563 \\
X-Learner & 1.1863 & 0.0286 & 0.0042 & 0.0563 \\
\rowcolor{gray!10}Causal BART & 0.9924 & 0.0165 & 0.0021 & 0.0523 \\ \hline
\end{tabular}
\caption{Simulation study for HTE estimation, with $M=2$ rules, no-confounder (randomized controlled trial), 2,000 individuals. For all the methods, the mean ($\mu$) and standard deviation ($\sigma$) treatment effect root mean squared error (RMSE) and bias (Bias) over 250 Monte Carlo experiments are reported.}
\label{tab:estimation_rct}
\end{table}
As discussed in section \ref{sec:sym_estimation}, (almost) all the CRE variants significantly outperform the corresponding `standalone' pseudo-outcome estimators in both pseudo-outcome and ATE estimation without significantly worsening the performances from the original data-generating process (with linear confounding), with exceptions of BCF. Given the similarity of the results with the original data-generating process, we empirically observe that under Assumption \eqref{ass:ignorability} CRE algorithm is robust with respect to the confounding mechanism. CRE (AIPW) is the best-performing method in pseudo-outcome estimation (although the unstable AIPW pseudo-outcome estimation) and CRE (Causal BART) leads to the most consistent estimate.

\begin{figure}[h!]
	\centering
	\includegraphics[width=\textwidth]{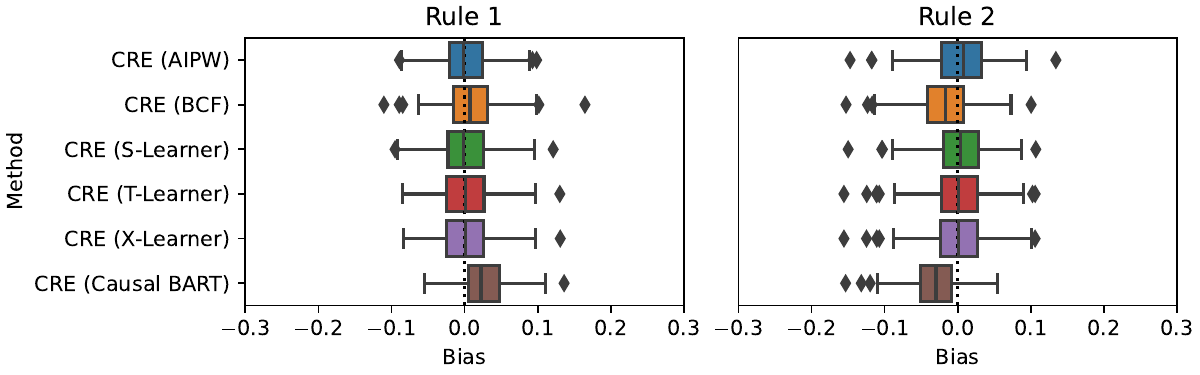} 
	\caption{Simulation study for HTE estimation, with $M=2$ rules, no-confounding (randomized controlled trial) and 2,000 individuals. For all the CRE variants, for each rule, the AATE's bias over 250 Monte Carlo experiments is reported in a boxplot.}
	\label{fig:alpha_rct}
\end{figure}

In Figure \ref{fig:alpha_rct}, we report the results for AATEs estimation, with no confounding (if not in terms of decision rules).
As expected from Proposition \ref{prop:alpha_asnormality}, all the methods lead to consistent AATEs estimation.

\subsubsection{Non-Linear Confounding}
In Table \ref{tab:estimation_nonlin_conf}, we report the results for heterogeneous treatment effect estimation with non-linear confounding.
\begin{table}[h!]
\centering
\begin{tabular}{crrrr}
\hline
 & \multicolumn{2}{c}{RMSE} & \multicolumn{2}{c}{Bias} \\
Method & \multicolumn{1}{c}{$\mu$} & \multicolumn{1}{c}{$\sigma$} & \multicolumn{1}{c}{$\mu$} & \multicolumn{1}{c}{$\sigma$} \\ \midrule \midrule
CRE (AIPW) & \textbf{0.1343} & 0.0601 & 0.0037 & 0.0881 \\
\rowcolor{gray!10}CRE (BCF) & 0.1493 & 0.0555 & 0.0050 & 0.0790 \\
CRE (S-Learner) & 0.1492 & 0.0599 & 0.0035 & 0.0849 \\
\rowcolor{gray!10}CRE (T-Learner) & 0.1490 & 0.0664 & 0.0074 & 0.0937 \\
CRE (X-Learner) & 0.1460 & 0.0651 & 0.0073 & 0.0938 \\
\rowcolor{gray!10}CRE (Causal BART) & 0.1421 & 0.0628 & \textbf{0.0002} & 0.0817 \\
\hline
AIPW & 2.0767 & 0.1945 & 0.0043 & 0.0560 \\
\rowcolor{gray!10}BCF & 0.1347 & 0.0370 & 0.0045 & 0.0524 \\
S-Learner & 0.4721 & 0.0333 & 0.0033 & 0.0529 \\
\rowcolor{gray!10}T-Learner & 0.8052 & 0.0373 & 0.0048 & 0.0568 \\
X-Learner & 1.1870 & 0.0292 & 0.0048 & 0.0568 \\
\rowcolor{gray!10}Causal BART & 0.9925 & 0.0164 & 0.0019 & 0.0517 \\ \hline
\end{tabular}
\caption{Simulation study for HTE estimation, with $M=2$ rules, non-linear confounder, 2,000 individuals. For all the methods, the mean ($\mu$) and standard deviation ($\sigma$) treatment effect root mean squared error (RMSE) and bias (Bias) over 250 Monte Carlo experiments are reported.}
\label{tab:estimation_nonlin_conf}
\end{table}
The confounding mechanism doesn't seem to significantly impact the estimation performances, and the results obtained look very similar to the ones from the original data-generating process and the randomized controlled experiment variant.
As discussed in section \ref{sec:sym_estimation}, (almost) all the CRE variants significantly outperform the corresponding `standalone' pseudo-outcome estimators in both pseudo-outcome and ATE estimation without significantly worsening the performances from the original data-generating process (with linear confounding), with exceptions of BCF.

\begin{figure}[h!]
	\centering
	\includegraphics[width=\textwidth]{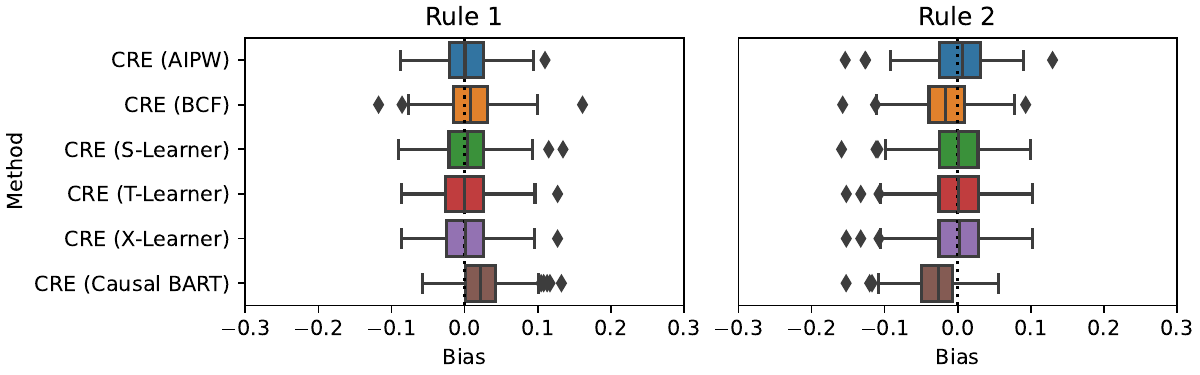} 
	\caption{Simulation study for HTE estimation, with $M=2$ rules, non-linear confounding and 2,000 individuals. For all the CRE variants, for each rule, the AATE's bias over 250 Monte Carlo experiments is reported in a boxplot.}
	\label{fig:alpha_nonlin_conf}
\end{figure}

In Figure \ref{fig:alpha_nonlin_conf}, we report the results for AATEs estimation, with non-linear confounding.
As expected from Proposition \ref{prop:alpha_asnormality}, all the methods lead to consistent AATEs estimation.

%% file: Appendix/appendix_d.tex
\section{Parameter Settings}
\label{app:parameters}

In this section, we summarize the parameter settings.
In Table \ref{tab:parameters}, we report the default Causal Rule Ensemble method and hyper-parameters used for the simulation studies relying on CRE 0.2.1 package \citeApp{khoshnevis2023}.
In Table \ref{tab:parameters_exp}, we report the default Causal Rule Ensemble method and hyperparameters used for the applied experiment on the discovery and inference of HTE of air pollution exposure on mortality relying on CRE 0.2.1 package \citeApp{khoshnevis2023}.. In both the tables, `XGboost' stands for the scalable end-to-end tree-boosting system algorithm by \citeApp{chen2016xgboost}.

\begin{table}[H]
\centering
\begin{tabular}{llc}
\midrule
 & Parameter & Value \\ \midrule  %\midrule
Honest Splitting & Ratio & 0.5 \\ 
&& \\
\textbf{Discovery} & \\ \midrule
\multirow{2}{*}{Pseudo-outcome Estimation} & Propensity Score estimator ($\hat{e}$) & XGboost \\
 & Outcome estimator ($\hat{\mu}$) & XGBoost \\
 \hline
\multicolumn{1}{c}{\multirow{6}{*}{Rules Generation}} & N. Trees (Random Forest) & 40 \\
\multicolumn{1}{c}{} & N. Trees (GBM) & 40 \\
\multicolumn{1}{c}{} & Replace & True \\
\multicolumn{1}{c}{} & Max node (subgroup) size & 20 \\
\multicolumn{1}{c}{} & Max depth (L) & 3 \\
\multicolumn{1}{c}{} & Max number of nodes (per tree) & 2$^3$ = 8 \\
\hline
\multirow{3}{*}{Filtering} & $t_{\text{decay}}$ & 0.025 \\
 & $t_{\text{ext}}$ & 0.01 \\
 & $t_{\text{corr}}$ & 1 \\
 \hline
\multirow{2}{*}{Rules Selection} & Upper Bound PFER & $\frac{L}{k+1}$ \\
 & $\pi_{\text{thr}}$ & 0.8 \\ 
\hline 
&& \\
\textbf{Estimation} & & \\ 
\midrule
\multirow{2}{*}{Pseudo-outcome Estimation} & Propensity Score estimator ($\hat{e}$) & XGboost \\
 & Outcome estimator ($\hat{\mu}$) & XGBoost \\ 
 \hline
 CATE Estimation & $t_{p-\text{value}}$ & 0.05 \\
 \midrule
\end{tabular}
\caption{List of CRE method parameters and hyperparameters used for the simulations.}
\label{tab:parameters}
\end{table}

\begin{table}[h!]
\centering
\begin{tabular}{llc}
\midrule
 & Parameter & Value \\ \midrule  %\midrule
Honest Splitting & Ratio & 0.5 \\ 
&& \\
\textbf{Discovery} & \\ \midrule
\multirow{2}{*}{Pseudo-outcome Estimation} & Estimator & X-Learner \\
 & Outcome estimator ($\hat{\mu}$) & XGBoost \\
 \hline
\multicolumn{1}{c}{\multirow{6}{*}{Rules Generation}} & N. Trees (Random Forest) & 100 \\
\multicolumn{1}{c}{} & N. Trees (GBM) & 100 \\
\multicolumn{1}{c}{} & Replace & True \\
\multicolumn{1}{c}{} & Max node (subgroup) size & 20 \\
\multicolumn{1}{c}{} & Max depth (L) & 2 \\
\multicolumn{1}{c}{} & Max number of nodes (per tree) & 4 \\
\hline
\multirow{3}{*}{Filtering} & $t_{\text{decay}}$ & 0.002 \\
 & $t_{\text{ext}}$ & 0.005 \\
 & $t_{\text{corr}}$ & 1 \\
 \hline
\multirow{2}{*}{Rules Selection} & Upper Bound PFER & 1 \\
 & $\pi_{\text{thr}}$ & 0.8 \\ 
\hline 
&& \\
\textbf{Estimation} & & \\ 
\midrule
\multirow{2}{*}{Pseudo-outcome Estimation} & Estimator & X-Learner \\
 & Outcome estimator ($\hat{\mu}$) & XGBoost \\ 
 \hline
 CATE Estimation & $t_{p-\text{value}}$ & 0.05 \\
 \midrule
\end{tabular}
\caption{List of CRE method parameters and hyper-parameters used for the discovery and estimation of HTE of air pollution exposure on mortality.}
\label{tab:parameters_exp}
\end{table}

%% file: Appendix/appendix_f.tex
\section{Testing for Spatial Auto-correlation}
\label{ssec:spatial_autocor}

To assess the robustness of the experimental results to spatial auto-correlation, we employed Moran's I \citepApp{moran1950test}. Specifically, we conducted tests to examine the presence of spatial auto-correlation in the residuals of the models used to impute the potential outcomes. Given that we possess information regarding the ZIP code of residence for each individual in our analysis, we explored potential auto-correlations between individuals within a particular ZIP code and individuals residing in neighboring ZIP codes. The results of these tests are presented in Table \ref{tab:Moran}. Notably, our findings indicate no statistical evidence of spatial auto-correlation across all four census regions. This suggests that spatial auto-correlation does not seem to influence our results significantly and reinforces the robustness of our analysis to spatial dependencies.

\begin{table}[]
\centering
\begin{tabular}{lcccc}
\hline
 & West & Midwest & South & Northeast \\
\midrule 
\midrule
Moran's I & 0.0110 & 0.0074 & 0.0112 & -0.0072 \\
p-value & 0.1310 & 0.1467 & 0.0595 & 0.7363 \\
\hline
\end{tabular}
\caption{Results of Moran's test for spatial auto-correlation in the four U.S. regions.}
\label{tab:Moran}
\end{table}

%% file: Appendix/appendix_g.tex
\section{Sensitivity Analysis}
\label{app:sensi}

The estimator $\hat{\bm{\alpha}}$ compactly represents the treatment effect heterogeneity. The validity and consistency of $\hat{\bm{\alpha}}$ rely on the assumption of no unmeasured confounders and correct specification of the propensity score model. However, in practice, we do not know whether a considered set $\bm{X}_i$ is sufficient for the unconfounded assumption to hold. When there exists a source of unmeasured confounding, this assumption is violated, and the identification results do not hold. Sensitivity analysis can be a useful tool to investigate the impact of unmeasured confounding bias. 
	
We propose a new sensitivity analysis method that can assess the robustness of our conclusion. This method does not attempt to quantify the degree of unmeasured confounding bias in a given data set. Instead, it is more realistic to see how our causal conclusion will change with respect to various degrees of such bias. 

We consider the marginal sensitivity model that was introduced by \citeApp{tan2006distributional} and \citeApp{zhao2019sensitivity}. Let the true propensity probability $e_0(\bx, y; z) = \pr_0 (Z=1 | \bX = \bx, Y(z)=y)$ for $z \in \{0, 1\}$. If the assumption of no unmeasured confounders holds, this probability would be the same as $e_0(\bx) = \pr_0 (Z=1 | \bX = \bx)$ (i.e., $e_0(\bx, y; z) = e_0(\bx)$) that is identifiable from the data. Unfortunately, this assumption cannot be tested since $e_0(\bx, y; z)$ is generally not identifiable from the data. In addition to this non-identifiability of $e_0(\bx, y; z)$, there is another difficulty in obtaining $e_0(\bx)$ non-parametrically when $\bX$ is high-dimensional. In practice, $e_0(\bx)$ is estimated by a parametric logistic model in the form of $e_{\gamma}(\bx) = \exp(\gamma' \bx)/\{1 + \exp(\gamma' \bx)\}$ where $e_{\gamma_0}(\bx)$  can be considered as the best parametric approximation of $e_0(\bx)$, and used for sensitivity analysis.
	
Our sensitivity model uses the sensitivity parameter $\Lambda \geq 1$ that restricts the maximum deviation of $e_0(\bx, y; z)$ from the identifiable quantity $e_{\gamma_0}(\bx) = \pr_{\gamma_0}(Z=1 | \bX = \bx)$. Sensitivity analysis is conducted for each $\Lambda$ to see if there is any qualitative change of our conclusion. More formally, define the set $\mathcal{E}_{\gamma_0}(\Lambda)$and assume $e_{0}(\bx) \in \mathcal{E}_{\gamma_0}(\Lambda)$,
\begin{equation}
		e_0(\bx, y; z) \in \mathcal{E}_{\gamma_0}(\Lambda) = \left\{ 0 < e(\bx, y; z) < 1: \frac{1}{\Lambda} \leq \frac{(1-e(\bx, y; z)) \cdot e_{\gamma_0}(\bx)}{e(\bx, y; z)\cdot(1- e_{\gamma_0}(\bx))} \leq \Lambda \right\} \quad \text{for  } z \in \{0, 1\}.
		\label{eqn:sensi_model}
	\end{equation}

The deviation of $e_0(\bx, y; z)$ is symmetric with respect to the parametrically identifiable quantity $e_{\gamma_0}(\bx)$, and the degree of the deviation is governed by the sensitivity parameter $\Lambda$. When $\Lambda = 1$, $e_0(\bx, y; z) = e_{\gamma_0}(\bx)$ for all $z$, which implies no violations of the following assumptions;  the correctly specified propensity score model assumption (that is, $e_{\gamma_0}(\bx) = e_0(\bx)$) and no unmeasured confounder assumption (that is, $e_0(\bx, y; z) = e_0(\bx)$). For $\Lambda>1$, the proposed sensitivity model considers violations of both assumptions. This model resembles the model proposed by \citeApp{rosenbaum2002observational}. The connection between the two models is illustrated in Section 7.1 in \citeApp{zhao2019sensitivity}.

%% file: Appendix/appendix_e.tex
\section{Additional Figures}

%\begin{figure}[h!]
% 	\centering	\includegraphics[width=\textwidth]{Images/map_pm25.jpeg} 
% 	\caption{Average levels PM$_{2.5}$ for the biennium 2010-2011 in the contiguous U.S.}
% 	\label{fig:map_pm25}
%\end{figure}

\begin{figure}[h!]
	\centering
	\includegraphics[width=\textwidth]{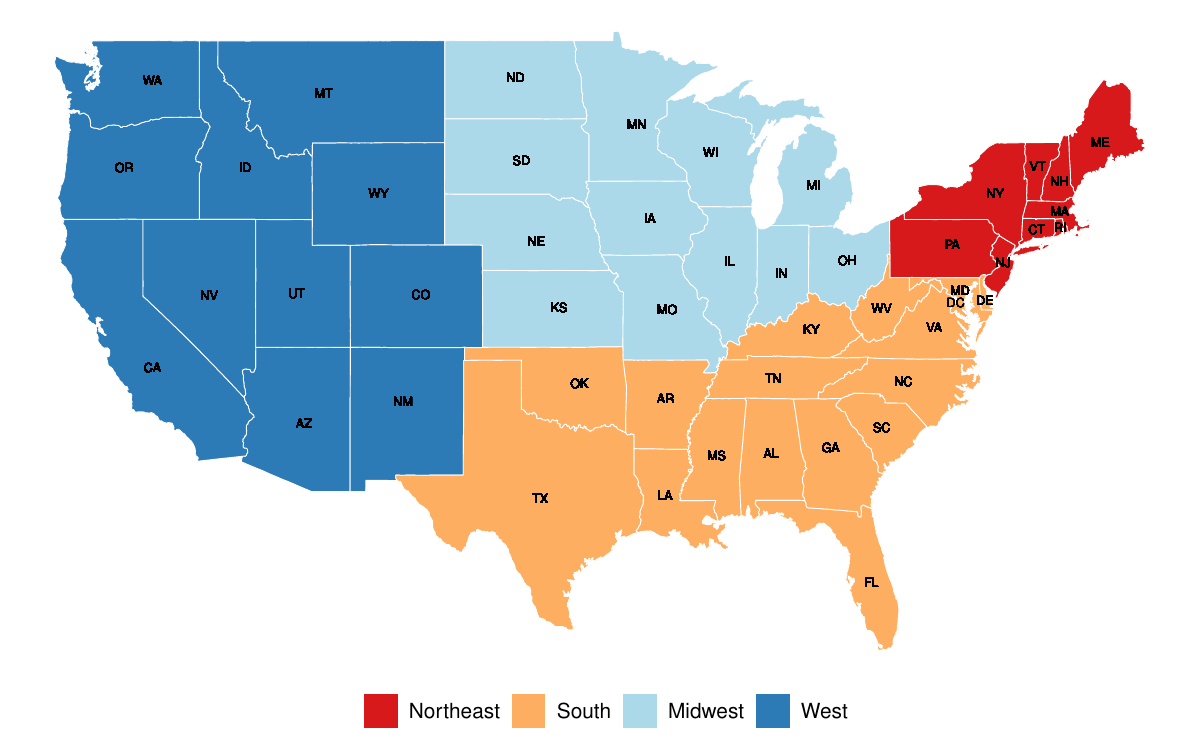} 
	\caption{Map of the four census geographical regions for the contiguous U.S.}
	\label{fig:regions}
\end{figure}